%% file: paper.tex
\documentclass{vldb}

\usepackage{algorithm}
\usepackage{algpseudocode}

\usepackage{graphicx}
\usepackage{balance}  
\usepackage{subfigure}
\usepackage{textcomp}
\usepackage{subfigure}
\usepackage{url}

\usepackage{color}

\begin{document}
%
\title{Bleach: A Distributed Stream Data Cleaning System}

\numberofauthors{3} 

\author{
\alignauthor
Yongchao Tian\\
       \affaddr{Eurecom\\
       yongchao.tian@eurecom.fr}
\alignauthor
Pietro Michiardi\\
       \affaddr{Eurecom\\
       pietro.michiardi@eurecom.fr}
\alignauthor
Marko Vukoli{\'c}\\
       \affaddr{IBM Research - Zurich\\
       mvu@zurich.ibm.com}
}

\maketitle

\begin{abstract}
\input{00-abstract}
\end{abstract}


%

\input{01-introduction}

\input{02-preliminary}

\input{03-design}

\input{04-dynamicrule}

\input{05-windowing}
\input{06-evaluation}
\input{07-relatedwork}
\input{08-futurework}



{
\bibliographystyle{abbrv}
\bibliography{paper}
}

\end{document}

%% file: 00-abstract.tex
In this paper we address the problem of rule-based stream data cleaning, which sets stringent requirements on latency, rule dynamics and ability to cope with the unbounded nature of data streams. 

We design a system, called Bleach, which achieves real-time violation detection and data repair on a dirty data stream. Bleach relies on efficient, compact and distributed data structures to maintain the necessary state to repair data, using an incremental version of the equivalence class algorithm. Additionally, it supports rule dynamics and uses a ``cumulative'' sliding window operation to improve cleaning accuracy.

We evaluate a prototype of Bleach using a TPC-DS derived dirty data stream and observe its high throughput, low latency and high cleaning accuracy, even with rule dynamics. Experimental results indicate superior performance of Bleach compared to a baseline system built on the micro-batch streaming paradigm.

%% file: 01-introduction.tex
\section{Introduction}
\label{sec:intro}

Today, we live in a world where decisions are often based on analytics applications that process continuous streams of data. 
Typically, data streams are combined and summarized to obtain a succint representation thereof: analytics applications rely on such representations to make predictions, and to create reports, dashboards and visualizations \cite{gdeltproject,recordedfuture,spark-summit}. All these applications expect the data, and their representation, to meet certain quality criteria. Data quality issues interfere with these representations and distort the data, leading to misleading analysis outcomes and potentially bad decisions.

As such, a range of data cleaning techniques were proposed recently~\cite{trifacta,openrefine,Khayyat:2015:BSB:2723372.2747646}. However, most of them focus on ``batch'' data cleaning, by processing static data stored in data warehouses, thus neglecting the important class of streaming data. In this paper, we address this gap and focus on \emph{stream data cleaning}. The challenge in stream cleaning is that it requires both \emph{real-time} guarantees as well as high \emph{accuracy}, requirements that are often at odds. 

A na{\"i}ve approach to stream data cleaning could simply extend existing batch techniques, by buffering data records in a temporary data store and cleaning it periodically before feeding it into downstream components. Although likely to achieve high accuracy, such a method clearly violates real-time requirements of streaming applications. 
The problem is exacerbated by the volume of data cleaning systems need to process, which prohibits centralized solutions. Therefore, our goal is to design a \emph{distributed stream data cleaning system}, which achieves efficient and accurate cleaning in real-time.

In this paper, we focus on rule-based data cleaning, whereby a set of domain-specific rules define how data should be cleaned: in particular, we consider functional dependencies (FDs) and conditional functional dependencies (CFDs).
Our system, called Bleach, proceeds in two phases: \emph{violation detection}, to find rule violations, and \emph{violation repair}, to repair data based on such violations. Bleach relies on efficient, compact and distributed data structures to maintain the necessary state (e.g., summaries of past data) to repair data, using an incremental equivalence class algorithm.

We further address the complications due to the long-term and dynamic nature of data streams: the definition of dirty data could change to follow such dynamics. Bleach supports dynamic rules, which can be added and deleted without requiring idle time. Additionally, Bleach implements a sliding window operation that trades modest additional storage requirements to temporarily store cumulative statistics, for increasing cleaning accuracy.

Our experimental performance evaluation of Bleach is two-fold. First, we study the performance, in terms of throughput, latency and accuracy, of our prototype and focus on the impact of its parameters. Then we compare Bleach to an alternative baseline system, which we implement using a micro-batch streaming architecture. Our results indicate the benefits of a system like Bleach, which hold even with rule dynamics.
Despite extensive work on rule-based data cleaning~\cite{abedjan2015temporal,4221723,6544847,Dallachiesa:2013:NCD:2463676.2465327,6243140,Kolahi:2009:AOR:1514894.1514901,DBLP:conf/dasfaa/ChenTHS015,Khayyat:2015:BSB:2723372.2747646}, we are not aware of any other stream data cleaning system.

The paper is organized as follows. Section~\ref{sec:pre} introduces our problem statement. The system design of Bleach is discussed in Section~\ref{sec:design}; dynamic rule management and windowing are discussed respectively in Section~\ref{sec:dyn} and Section~\ref{sec:window}. Section~\ref{sec:eva} presents our experimental results. Section~\ref{sec:rel} overviews related work. Finally, Section~\ref{sec:con} concludes our work.

%% file: 02-preliminary.tex
\section{Preliminaries}
\label{sec:pre}
Next, we introduce some basic notations we use throughout the paper, then we define the problem statement we consider in this work.

\subsection{Background and Definitions}

Similar to data cleaning systems in data warehouses, that read a dirty dataset and write back a cleaned dataset, in this paper we assume that a stream data cleaning system ingests a data stream and outputs a cleaned data stream instance.
We consider an input data stream instance $D_{in}$ with schema $S(A_1, A_2, ..., A_m)$ where $A_j$ is an attribute in schema $S$. 
We assume the existence of unique tuple identifiers for every tuple in $D_{in}$: 
thus given a tuple $t_i$, $id(t_i)$ is $t_i$'s identifier.
In general we define a function $id(e)$ which returns the identifier (ID) of $e$ where $e$ can be any element.
A list of IDs $[id(e_1),id(e_2),...,id(e_n)]$ is expressed as $id(e_1,e_2,...,e_n)$ for brevity.
The output data stream instance $D_{out}$ complies to schema $S$ and has the same tuple identifiers as in $D_{in}$, without any loss or duplication.
The basic unit, a \emph{cell} $c_{i,j}$, 
is the concatenation of a tuple id, an attribute and the projection of the tuple on the attribute:
$c_{i,j}=(id(t_i),A_j,t_i(A_j))$.
Note that $t_i(A_j)$ is the value of $c_{i,j}$, which can also be expressed as $v(c_{i,j})$.
Sometimes, we may simply express $c_{i,j}$ as $c_i$ when the cell attribute is not relevant to the discussion.
In our work, when we point at a specific tuple $t_i$, we also refer to this tuple as the \emph{current} tuple.
Tuples appearing earlier than $t_i$ in the data stream are referred to as \emph{earlier} tuples and those appearing after $t_i$ are referred to as \emph{later} tuples.

To perform data cleaning, we define a set of rules $\Sigma = [r_1, ..., r_n ]$, in which $r_k$ is either a functional dependency (FD) rule or a conditional FD rule (CFD). 
Each rule has a unique rule identifier $id(r_k)$.
A CFD rule $r_k$ is represented by $(X \rightarrow A, cond(Y))$, in which $cond(Y)$ is a boolean function on a set of attributes $Y$ where $Y \subseteq S$. 
$X$ and $A$ are respectively referred to as a set of left-hand side (LHS) attributes and right-hand side (RHS) attribute:
$LHS(r_k) = X$, $RHS(r_k) = A$. 
When the rule is clear in the context, we omit $r_k$ so that $LHS = X$, $RHS = A$.
Cells of LHS (RHS) attributes are also referred to as LHS (RHS) cells.
$Y$ is referred to as a set of conditional attributes.
For a pair of tuples $t_1$ and $t_2$ satisfying condition $cond(t_1(Y))=cond(t_2(Y))=true$, if $t_1(B) = t_2(B)$ for all $B \in X$ but $t_1(A) \neq t_2(A)$,
then it is a \emph{violation} for $r_k$. 
A data stream instance $D$ satisfies $r_k$, denoted as $D \models r_k$, when there are no violations for $r_k$ exist in $D$.
A FD rule can be seen as a special case of CFD rule where $cond(Y)$ is always true and $Y$ is $\emptyset$. 
We refer to an attribute as an \emph{intersecting} attribute if it is involved in multiple rules.

If $D$ satisfies a set of rules $\Sigma$, denoted $D \models \Sigma$, then $D \models r_k$ for $\forall r_k \in \Sigma$. If $D$ does not satisfy $\Sigma$, $D$ is a dirty data stream instance.

\subsection{Challenges and Goals}

An ideal stream data cleaning system should accept a dirty input stream $D_{in}$ and output a clean stream $D_{out}$, in which all \emph{rule violations} in $D_{in}$ are repaired ($D_{out} \models \Sigma$). However, this is not possible in reality
due to:

\begin{itemize}

\item \textbf{Real-time constraint}: As the data cleaning is incremental, 
the cleaning decision for a tuple (repair or not repair) can only be made based on itself and earlier tuples in the data stream, which is different from data cleaning in data warehouses where the entire dataset is available. 
In other words, if a dirty tuple only has violations with later tuples in the data stream, it can not be cleaned. A late update for a tuple in the output data stream is not accepted.

\item \textbf{Dynamic rules}: In a stream data cleaning system, the rule set is not static. A new rule may be added or an obsolete rule may be deleted at any time. A processed data tuple can not be cleaned again with an updated rule set. Reprocessing the whole data stream whenever the rule set is updated is not realistic.

\item \textbf{Unbounded data}: A data stream produces an unbounded amount of data, that cannot be stored completely. Thus, stream data cleaning can not afford to perform cleaning on the full data history. Namely, if a dirty tuple only has violations with tuples that appear much earlier in the data stream, it is likely that it will not be cleaned.

\end{itemize}

\begin{figure}
	\centering
	{\includegraphics[width=0.95\linewidth]{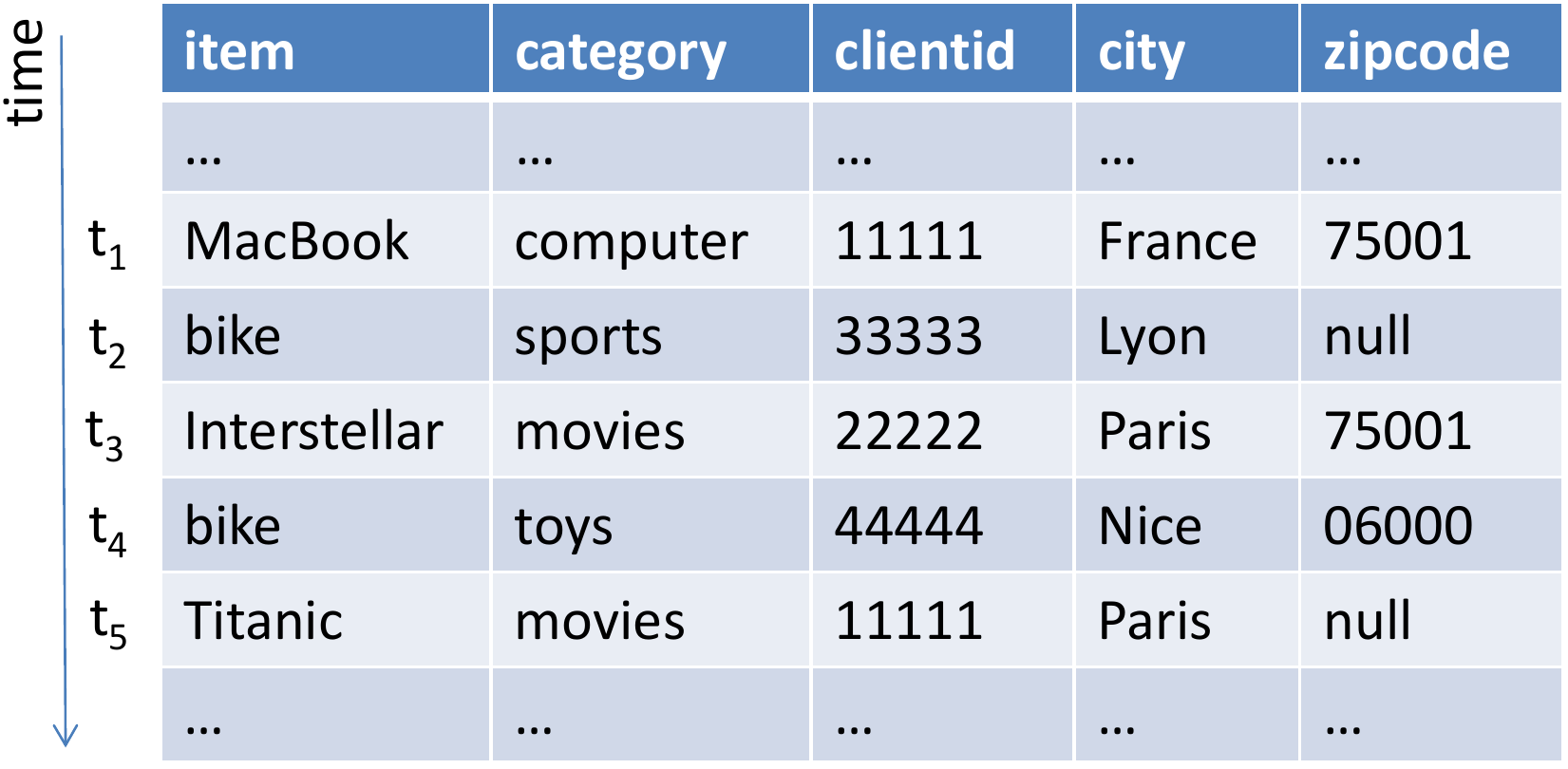}}
	\caption{Illustrative example of a data stream consisting of on-line transactions.}
	\label{fig:dataexample}
\end{figure}

Consider the example in Figure~\ref{fig:dataexample}, which is a data stream of on-line shopping transactions. Each tuple represents a purchase record, which contains a purchased item ($item$), the category of that item ($category$), a client identifier ($clientid$), the city of the client ($city$) and the zip code of that city ($zipcode$). In the example, we show an extract of five data tuples of the data stream, from $t_1$ to $t_5$.

Now, assume we are given two FD rules and one CFD rule stating how a clean data stream should look like: ($r1$) the same items can only belong to the same category; ($r2$) two records with the same $clientid$ must have the same city; ($r3$) two records with the same non-null zip code must have the same city:

\begin{description}
\item[($r_1$)] $item \rightarrow category$
\item[($r_2$)] $clientid \rightarrow city$
\item[($r_3$)] $zipcode \rightarrow city, zipcode \neq null$
\end{description}

If we focus on the detection of tuples that \emph{violate} such rules, we recognize three violations among the five tuples: ($v1$) $t_1$ and $t_3$ have the same non-null zip code ($t_1(zipcode)=t_2(zipcode) \neq null$) but different city names ($t_1(city) \neq t_2(city)$); ($v2$) $t_2$ claims bikes belong to category sports while $t_4$ classifies bikes as toys ($t_2(item)=t_4(item), t_2(category) \neq t_{4}(category)$); and ($v3$) $t_1$ and $t_5$ have the same $clientid$ but different city names ($t_1(clientid)=t_5(clientid), t_1(city) \neq t_5(city)$).

Note that when a stream data cleaning system receives tuple $t_1$, no violation can be detected as in our example $t_1$ only has violations with later tuples $t_3$ and $t_5$. Thus, no modification can be made to $t_1$. Furthermore, delaying the cleaning process for $t_1$ is not an option, not only because of real-time constraints, but also because it is difficult to predict for how long this tuple should be buffered for it to be cleaned.

Although performing incremental violation detection seems straightforward, incremental violation repair is much more complex to achieve.
Coming back to the example in Figure~\ref{fig:dataexample}, 
assume that the stream cleaning system receives tuple $t_5$ and successfully detects the violation $v_3$ between $t_5$ and $t_1$. Such detection is not sufficient to make the correct repair decision, as the tuple $t_1$ also conflicts with another tuple, $t_3$. An incremental repair in stream data cleaning system 
should also take the violations among earlier tuples into account.

\begin{figure}
	\centering
	{\includegraphics[width=0.75\linewidth]{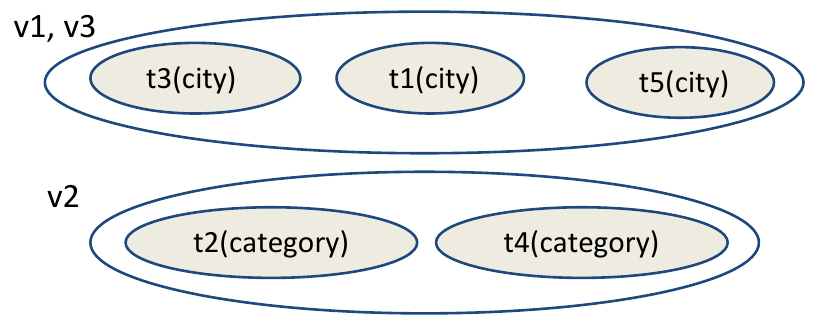}}
	\caption{An example of a violation graph, derived from our running example.}
	\label{fig:violationgraphexample}
\end{figure}

To account for the intricacies of the violation repair process, we introduce the concept of {\it violation graph}~\cite{Khayyat:2015:BSB:2723372.2747646}.
A violation graph is a data structure containing the detected violations, in which each node represents a cell.
If some violations share a common cell, they will be grouped into a single \emph{subgraph} (sg). Therefore, the violation graph is partitioned into smaller independent subgraphs. 
A single cell can only be in one subgraph. If two subgraphs share a common cell, they need to merge.

The repair decision of a tuple is only relevant to the subgraphs in which its cells are involved. A violation graph for our example can be seen in Figure~\ref{fig:violationgraphexample}. Given this violation graph, to make the repair decision for tuple $t_5$, the cleaning system can only rely on the upper subgraph which consists of violation $v_1$ and $v_3$ with the common cell $t_1(city)$.

We now give our problem statement as following.

\noindent \textbf{Problem statement}: Given an unbounded data stream with an associated schema\footnote{Note that although we restrict the data stream to have a fixed schema in this work, it is easy to extend our work to support a dynamic schema.} and a \emph{dynamic} set of rules, how can we design an \emph{incremental} and \emph{real-time} data cleaning system, including violation detection and violation repair mechanisms, using bounded computing and storage resources, to output a cleaned data stream?
In the next three sections, we give a detailed description of our distributed stream data cleaning system, that we call Bleach.

%% file: 03-design.tex
\section{Bleach Design and Algorithms}
\label{sec:design}

\begin{figure}
	\centering
	{\includegraphics[width=0.95\linewidth]{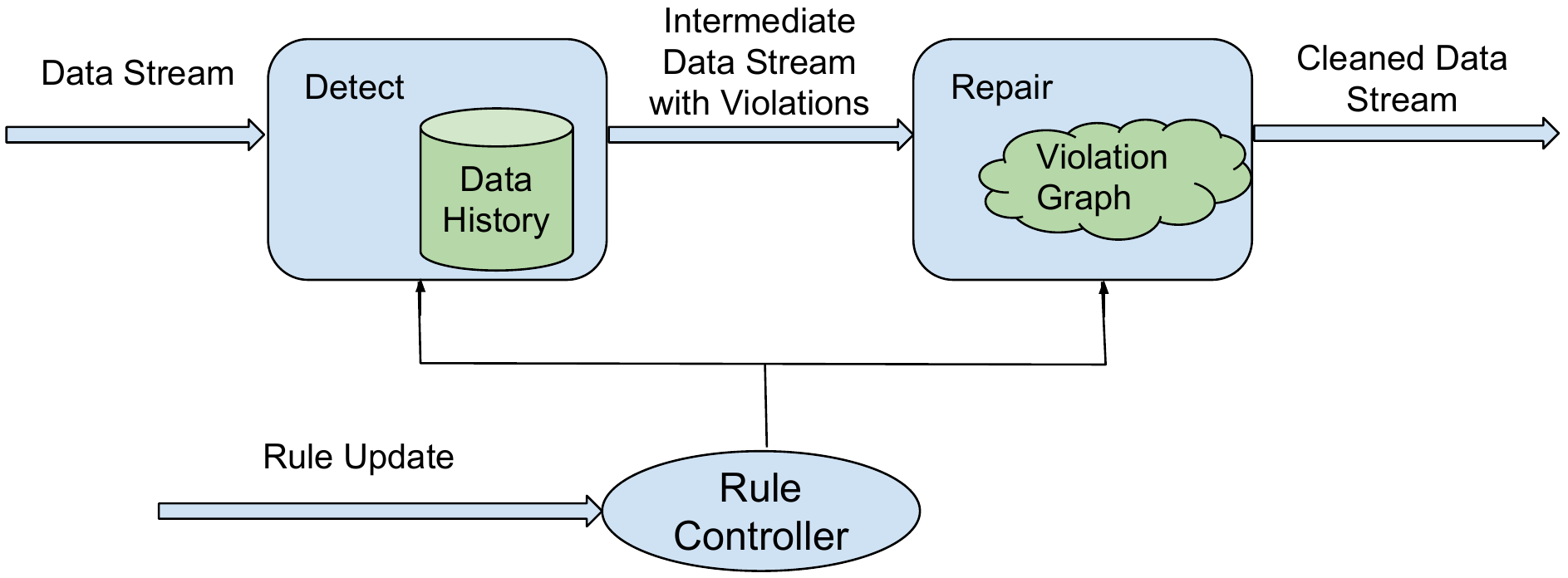}}
	\caption{Stream data cleaning Overview}
	\label{fig:overview}
\end{figure}

In this section, we overview the Bleach architecture and provide details about its components. As shown in Figure~\ref{fig:overview}, Bleach consists in two main blocks, namely the \emph{detect} and \emph{repair} modules, and a rule controller module, which is discussed in Section~\ref{sec:dyn}.

The input data stream first enters the detect module, which reveals violations against defined rules. The intermediate data stream, output from the first module, is enriched with violation information, which the repair module uses to make repair decisions. Finally, the system outputs a cleaned data stream.

Next, we delve into the details of the first two modules and outline several optimizations that aim at achieving efficiency and performance.

\subsection{Violation Detection}
\label{ssec:detect}

The violation detection module aims at finding input tuples that violate rules. To do so, it stores them in an in-memory, efficient and compact data structure that we call the \emph{data history}. Input tuples are thus compared to those in the data history to detect violations.
\begin{figure}
	\centering
	{\includegraphics[width=1\linewidth]{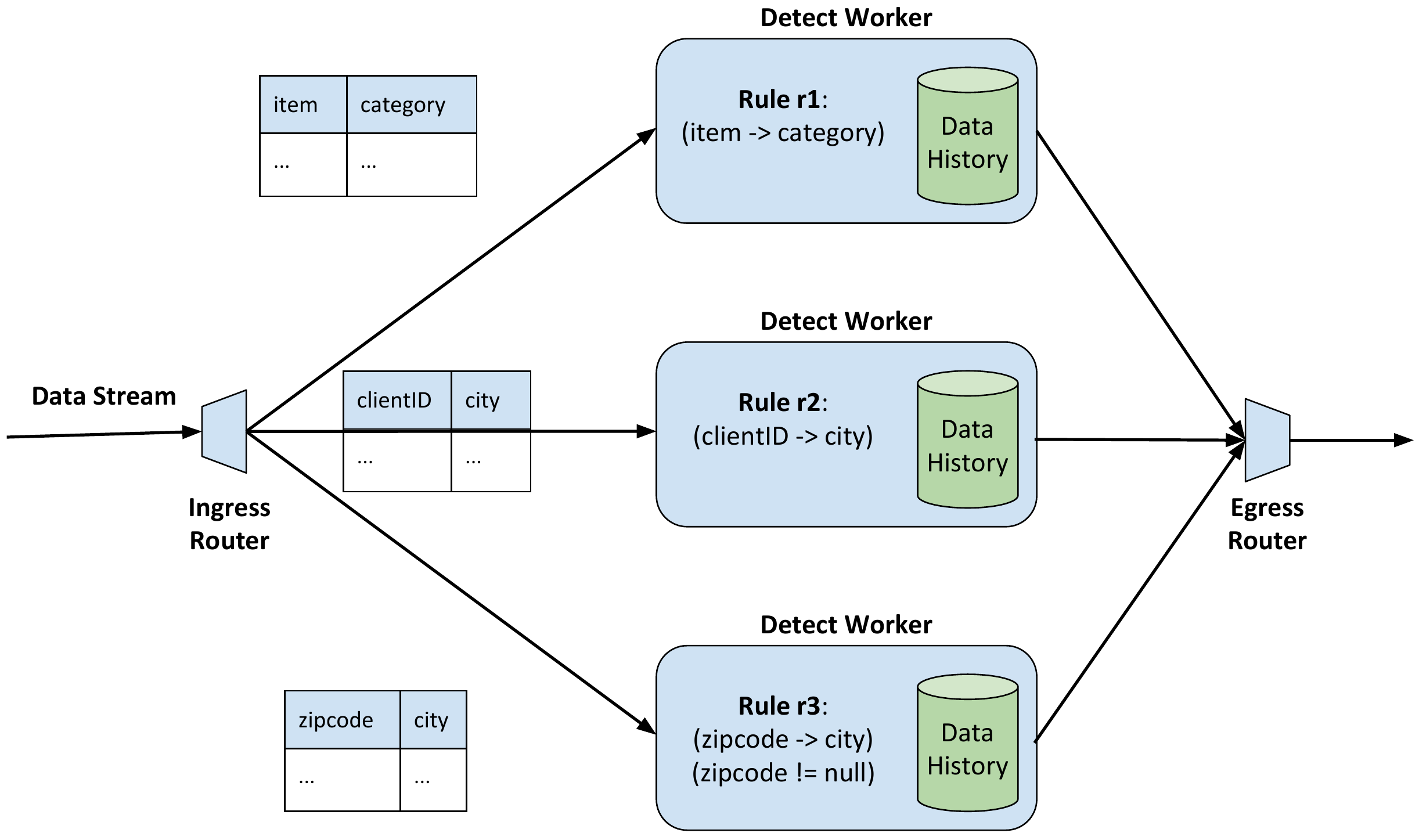}}
	\caption{The Detect Module}
	\label{fig:detect}
\end{figure}

Figure~\ref{fig:detect} illustrates the internals of the detect module: it consists of an ingress router, an egress router and multiple detect workers (DW). Bleach maps violation rules to such DW: each worker is in charge of finding violations for a specific rule.

\subsubsection{The Ingress Router}

The goal of the ingress router is to partition and distribute incoming tuples to DWs. Now, as discussed in Section \ref{sec:pre}, only a subset of the attributes of an input tuple are relevant when verifying data validity against a given rule. For example, a FD rule only requires its LHS and RHS attributes to be verified, ignoring the rest of the input tuple attributes.

Therefore, when the ingress router receives an input tuple, it partitions the tuple based on the current rule set, and only sends the relevant information to each DW in charge of each specific rule. As such, an input tuple is broken into multiple sub-tuples, which all share the same identifier of the corresponding input tuple. Note that some attributes of an input tuple might be required by multiple rules: in this case, sub-tuples will contain redundant information, allowing each DW to work independently. 

An example of tuple partitioning can be found in Figure~\ref{fig:detect}, where we reuse the input data schema and the rules from Section~\ref{sec:pre}.

\subsubsection{The Detect Worker}
\label{ssec:tdw}
Each DW is assigned a rule, and receives the relevant sub-tuples stemming from the input stream. For each sub-tuple, a DW needs to perform a lookup operation in the data history, and eventually emit a message (that is part of an intermediate data stream) to downstream components when a rule violation is detected. 

To achieve efficiency and performance, lookup operations on the data history need to be fast, and the intermediate data stream should avoid redundant information. Next, we first describe how the data history is represented and materializes in memory; then, we describe the output messages a DW generates, and finally outline the DW algorithm.

\noindent \textbf{Data history representation.} A DW accumulates relevant input sub-tuples in a compact data structure that enables an efficient lookup process, which makes it similar to a traditional indexing mechanism.

\begin{figure}
	\centering
	{\includegraphics[width=0.70\linewidth]{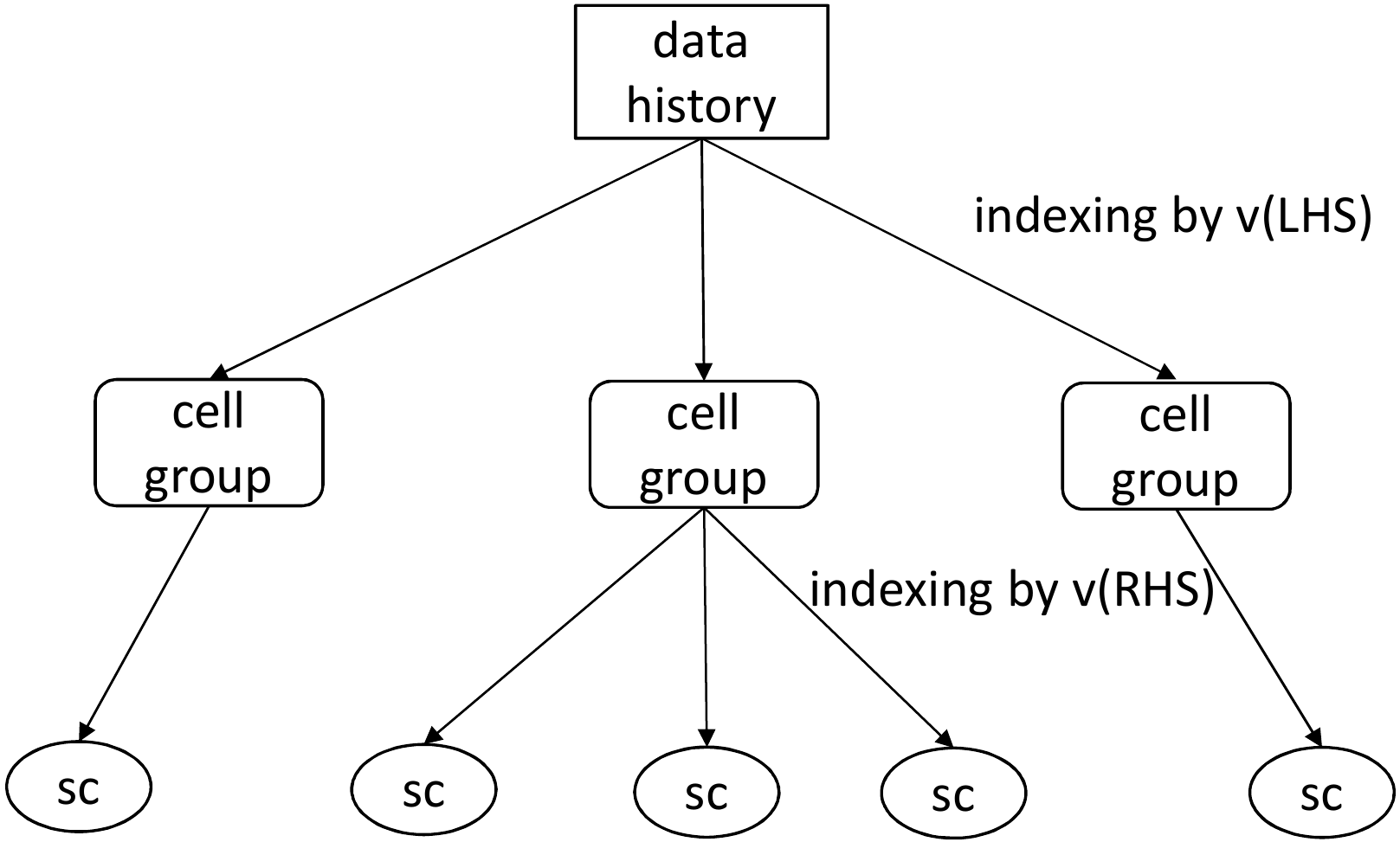}}
	\caption{The structure of the data history in a detect worker}
	\label{fig:datahistory}
\end{figure}

The structure\footnote{The techniques we use are similar to the notion of \emph{partitions} and \emph{compression} introduced in Nadeef~\cite{Dallachiesa:2013:NCD:2463676.2465327}.} of the data history is illustrated in Figure~\ref{fig:datahistory}. 
First, to speed-up the lookup process, sub-tuples are grouped by the value of 
the LHS attribute
used by a given rule: we call such group a \emph{cell group} (cg). 
Thus, A cg stores all RHS cells whose sub-tuples share the same LHS value. 
The identifier of a cell group $cg_l$ is
the combination of the rule assigned to the DW, and the value of LHS attributes, expressed as
$id(cg_l)=(id(r_k), t(LHS))$ where $r_k$ is the rule assigned to the DW. 

Next, to achieve a compact data representation, all cells in a cg sharing the same RHS value are grouped into a \emph{super cell} (sc): 
$sc_m=[c_{1,j}, c_{2,j},..., c_{n,j}]$.
From Section~\ref{sec:pre}, recall that a cell is made of a tuple ID, an attribute and a value: $(id(t_i), A_j, t_i(A_j))$. 
Therefore, a super cell can be \emph{compressed} as a list of tuple IDs, an attribute and their common value: 
$sc_m=(id(t_1,t_2,...,t_n), A_j, t(A_j))$ where $t(A_j)=t_1(A_j)=...=t_n(A_j)$.
Hence, within an individual DW, sub-tuples whose cells are compressed in the same \emph{sc} are equivalent, as they have the same LHS attributes value (the identity of the cell group) and the same RHS attribute value (the value of super cell). 
A cell group $cg_l$ now can be expressed as: 
$cg_l = ((id(r_k), t(LHS)), [sc_{1},sc_{2},...])$ including a identifier and a list of super cells.

In summary, the lookup process for a given input sub-tuple is as follows. Cell groups are stored in a hash-map using their identifier as keys: therefore the DW first finds the cg corresponding to the current sub-tuple. Cells in the corresponding cg are the only cells that  
might be in conflict with the current cell.
Overall, the complexity of the lookup process for a sub-tuple is $O(1)$.

\noindent \textbf{Violation messages.} DWs generate an intermediate data stream of \emph{violation messages}, which help downstream components to eventually repair input tuples. The goal of the DW is to generate as few messages as possible, while allowing effective data repair.

When the lookup process reveals the current tuple does not violate a rule, DWs emit a non-violation message ($msg_{nvio}$). Instead, when a violation is detected, a DW constructs a message with all the necessary information to repair it, including: the ID of the cell group corresponding to the current tuple and the RHS cells of the current and earlier tuples in data history:  
$msg_{vio} = (id(cg_l), c_{cur}, c_{old})$.

Now, to reduce the number of violation messages, the DW can use a super cell in place of a single cell ($c_{old}$) in conflict with the current tuple. In addition, recall that a single cg can contain multiple super cells, thus possibly requiring multiple messages for each group. However, we observe that two cells in the same cg must also conflict with each other, as long as their values are different. Since the data repair module in Bleach is stateful, it is safe to omit multiple violation messages for such cells.

\noindent \textbf{Algorithm details.} Next, we present the DW violation algorithm details, as illustrated in Algorithm~\ref{alg:viodetect}.

\begin{algorithm}
\caption{Violation Detection} \label{alg:viodetect}
\begin{algorithmic}[1]
\State given rule $r = (X \rightarrow A_j, cond(Y))$ 
\Procedure {Receive}{sub-tuple $t_i$} \Comment{$cond(t_i(Y))=true$}
\If {$\exists id(cg_l)=(id(r),t_i(X))$}
\If {$|cg_l| = 1$} \Comment{$cg_l$ contains $sc_{old}$}
\If {$v(sc_{old})=t_i(A_j)$}
\State Emit $msg_{nvio}$
\Else
\State Emit $msg_{vio}$ $(id(cg_l), c_{cur}, sc_{old})$
\EndIf
\Else
\State Emit $msg_{vio}$ $(id(cg_l), c_{cur}, null)$
\EndIf
\Else
\State Create $cg_l$ \Comment{Create a new cell group}
\State Emit $msg_{nvio}$
\EndIf
\State Add $c_{cur}$ to $cg_l$
\EndProcedure
\end{algorithmic}
\end{algorithm}

The algorithm starts by treating FD rules as a special case of CFD rules (line 1). 

Then, when a DW receives a sub-tuple $t_i$ satisfying the rule condition (line 2), it performs a lookup in the data history to check if the corresponding cell group $cg_l$ exists (line 3). If yes, it determines the number of \emph{sc} contained in the $cg_l$ (line 4). 

If there is only one sc $sc_{old}$, violation detection works as follows. If the RHS cell of the current sub-tuple, $c_{cur}$, has the same value as $sc_{old}$, it emits a non-violation message (line 5-6). Otherwise, a violation has been detected: the DW emits a \emph{complete} violation message, containing both the current cell and the old cell (line 8). 

If the cg contains more than one sc, 
the DW emits a single \emph{append-only} violation message, which only contains the cell of the current sub-tuple (line 11). Such compact messages omits the sc from the data history, since they must be contained in earlier violation messages.

Finally, if the lookup procedure (line 3) fails, the DW creates a new cell group and emits a non-violation message (line 14-15). 

At this point, the current cell $c_{cur}$ is added to the corresponding group $cg_l$ (line 17), either in an existing sc, or as a new distinct cell.

It is worth noticing that, following Algorithm~\ref{alg:viodetect}, a DW emits a single message for each input sub-tuple, no matter how many tuples in the data history it conflicts with.
 
\subsubsection{The Egress Router} 

The egress router gathers (violation or non-violation) messages for a given data tuple, as received from all DWs. Such messages are 
then sent together downstream towards the repair module.

\subsection{Violation Repair}
\label{ssec:repair}
The goal of this module is to take the repair decisions for dirty data tuples, based on an intermediate stream of violation messages generated by the detect module. 
To achieve data repair, Bleach uses a data structure called \emph{violation graph}, as outlined in Section~\ref{sec:pre}. 
Violation messages contribute to the creation and dynamics of the violation graph, which essentially groups those cells that, together, are used to perform data repair.

\begin{figure}
        \centering
        {\includegraphics[width=1\linewidth]{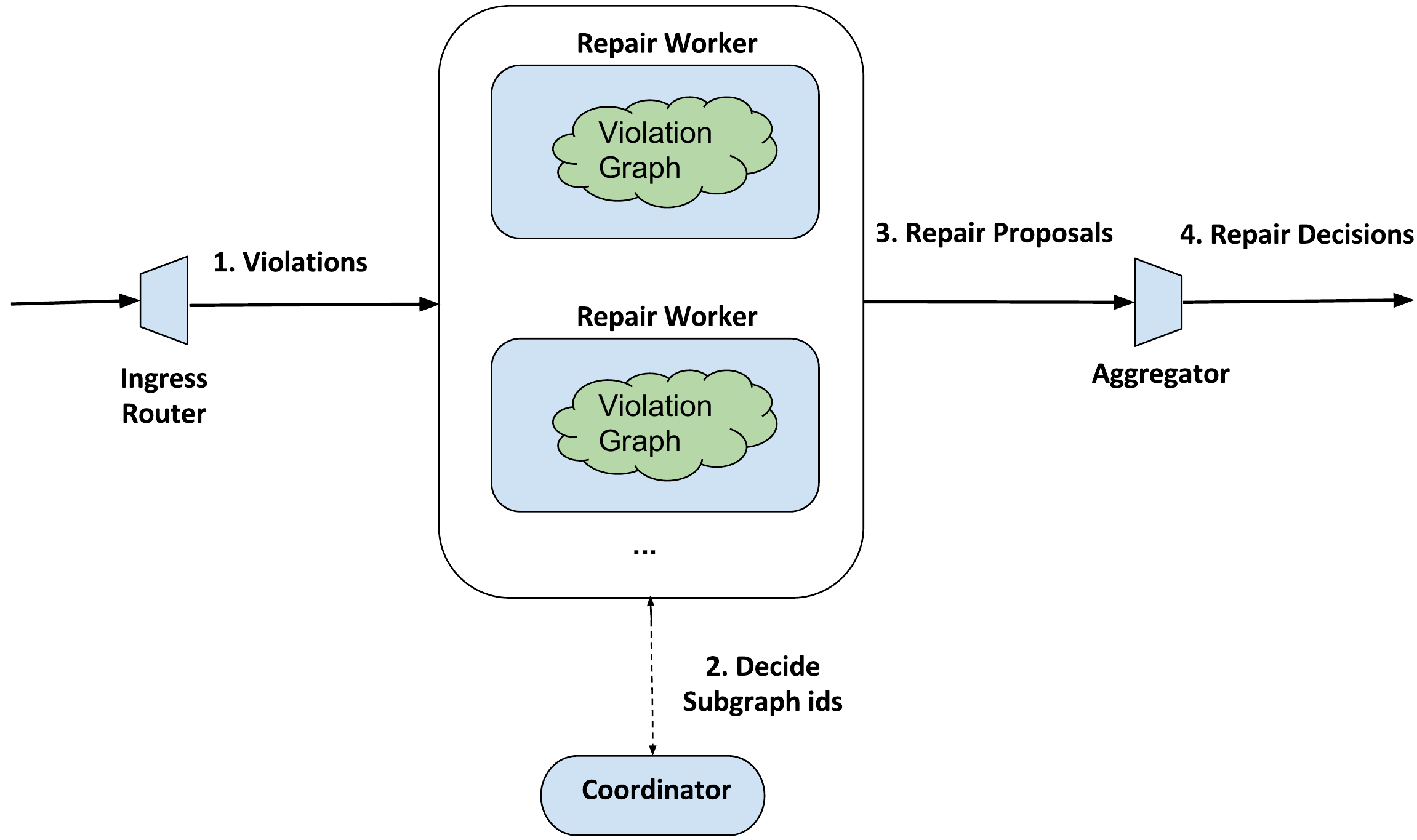}}
        \caption{Violation Repair}
        \label{fig:repair}
\end{figure}

Figure \ref{fig:repair} sketches the internals of the repair module: it consists of an ingress router, the repair workers (RW), and an aggregation component that emits clean data. An additional component, called the coordinator, is used to steer violation graph management, with the contribution of RWs.

\subsubsection{The Ingress Router}
The ingress router is a simple component that actually broadcasts all incoming violation messages to all RWs. As opposed to its counterpart in the detection module, the ingress router does not perform data partitioning: instead, RWs are in charge of using only relevant information contained in the violation messages they receive, with the goal of creating and maintaining the violation graph.

\subsubsection{The Repair Worker}
Next, we delve into the details of the operation of a RW. First, we focus on the violation graph and the data repair algorithm we implement in Bleach. Then, we move to the key challenge that RWs address, that is how to maintain a \emph{distributed} violation graph. As such, we focus on graph partitioning and maintenance. Due to violation graph dynamics, coordination issues might arise in a distributed setting: such problems are addressed by the coordinator component.

\noindent \textbf{The repair algorithm.} Current data repair algorithms use the concept of a violation graph to repair dirty data based on user-defined rules. As outlined in Section~\ref{sec:pre}, a violation graph is a succinct representation of cells (both current and historical) that are in conflict according to some rules. A violation graph is composed of subgraphs. 
As incoming data streams in, the violation graph evolves: specifically, its subgraphs might merge or split, depending on the contents of violation messages.

Using the violation graph, several algorithms can perform data cleaning, such as the equivalence class algorithm~\cite{Bohannon:2005:CME:1066157.1066175} or the holistic data cleaning algorithm~\cite{6544847}. Currently, Bleach relies on an \emph{incremental} version of the equivalence class algorithm, that supports streaming input data, although alternative approaches can be easily plugged in our system. Thus, a subgraph in the violation graph can be interpreted as an equivalence class, in which all cells are supposed to have the same value.

\noindent \textbf{Distributed violation graph.} 
Due to the unbounded nature of streaming data, it is reasonable to expect the violation graph to grow to sizes exceeding the capacity of a single RW. As such, in Bleach, the violation graph is a distributed data structure, partitioned across all RWs. 

However, unlike for DWs, the partitioning scheme can not be simply rule based, because a cell may violate multiple rules, creating issues related to coordination and load balancing. More generally, no partitioning scheme can guarantee that cells from a single violation message to be placed in the same partition
to store subgraphs in a single RW.

Next, we describe how Bleach builds and maintains a distributed violation graph. The graph is built using 
$msg_{vio}$
output by the egress router, which all RWs receive. 
Upon receiving a violation message, RWs process it independently, according to the following rules: {\it i)} if any of the current or old cells encapsulated in the message are already contained in an existing subgraph, both cells are added to the existing subgraph; 
{\it ii)} if an existing subgraph has cells which are in the same cell group as any of the cells in the message, the cells in the message are both added to the existing subgraph;
{\it iii)} if any of these two cells are contained in multiple subgraphs, these subgraphs need to merge;
{\it iv)} if none of these two cells is already contained in any subgraph, a new subgraph will be created with these two cells.

We define a \emph{subgraph identifier} $id(sg_k)$ to be the list of cell group IDs comprised in 
$msg_{vio}$:
$id(cg_1,cg_2,...)$. 
A subgraph can be expressed as 
$sg_k = (id(cg_1,cg_2,...), [sc_1, sc_2, ...])$:
it consists of a group of sc, stored in compressed format, as shown in Section~\ref{ssec:tdw}.
Note that when two subgraphs merge, their identifiers are also merged by concatenating both \emph{cg} ID lists.
To make the subgraph ID clear, $sg_k$ can be presented as $sg_{id(cg_1,cg_2,...)}$.

Since subgraphs are collections of cells, we distribute the latter across all RWs, using the cells tuple IDs for partitioning. Then, we use the subgraph identifier to recognize partitions from the same subgraph. As a consequence, a subgraph spans several RWs, each storing a fraction of the cells comprised in the subgraph.

Finally, we note that the violation graph, and in particular the subgraph partitions stored by each RW, materializes as a data structure stored in RAM. Such data structure is organized similarly to that of the data history presented in Section~\ref{ssec:detect}, which allows an efficient execution of the repair algorithm, and a compact data representation.

\begin{figure}[]
\centering
\subfigure[initial state]{
\includegraphics[width=0.45\linewidth]{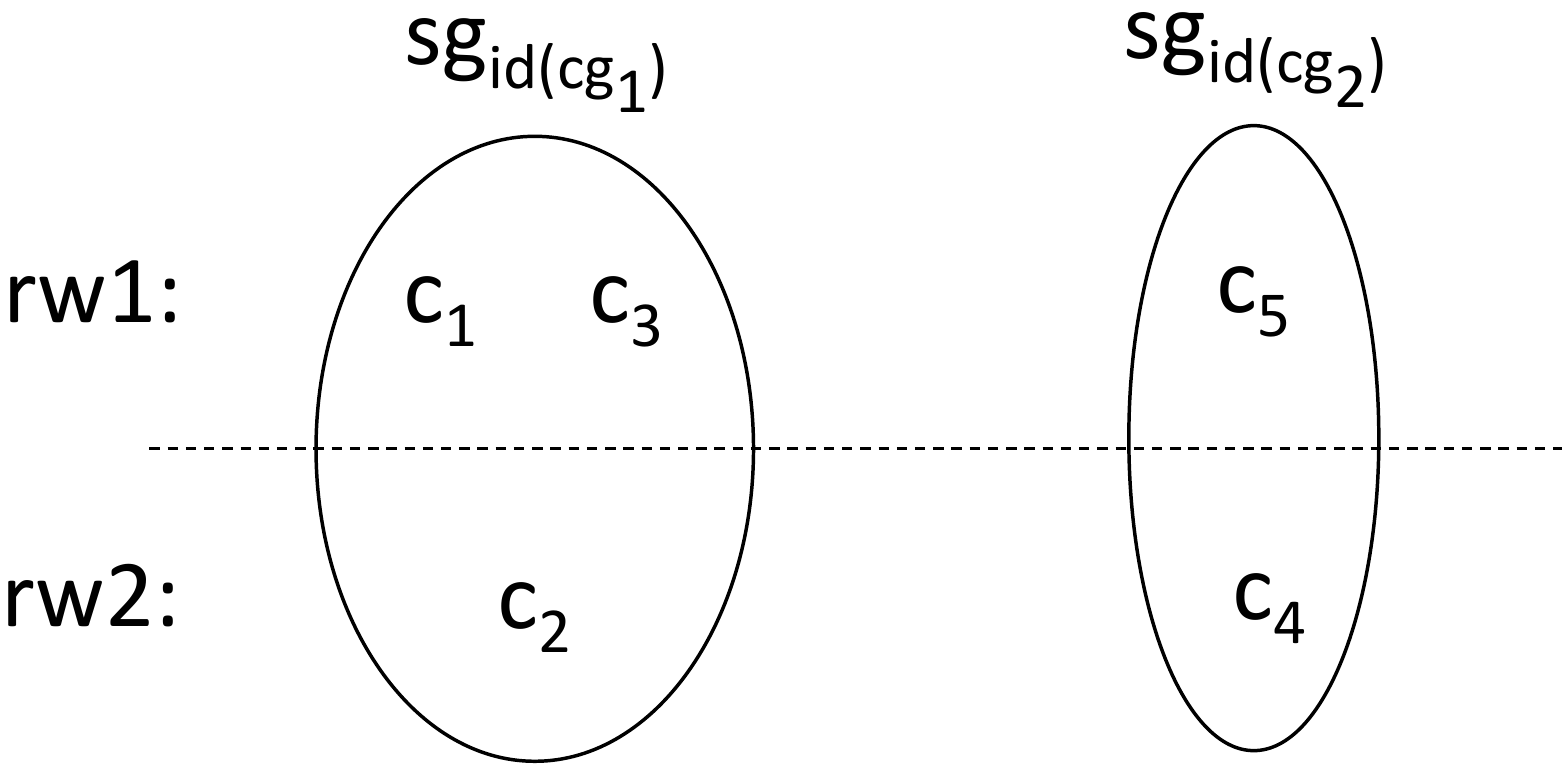}
\label{fig:subgraph1}}
\quad
\subfigure[merge only in $rw1$]{%
\includegraphics[width=0.45\linewidth]{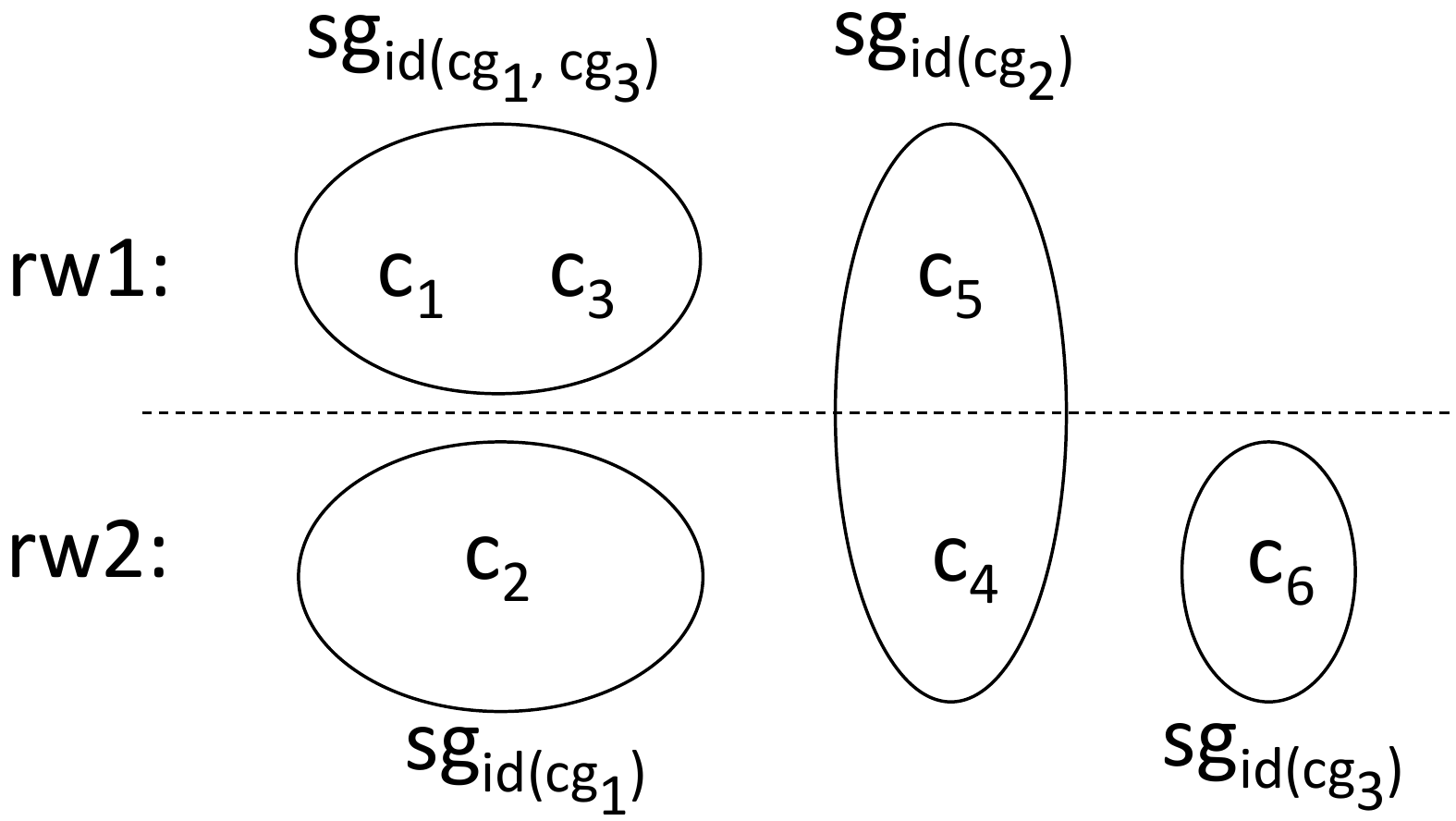}
\label{fig:subgraph3}}
\subfigure[merge in $rw1$ and $rw2$]{%
\includegraphics[width=0.45\linewidth]{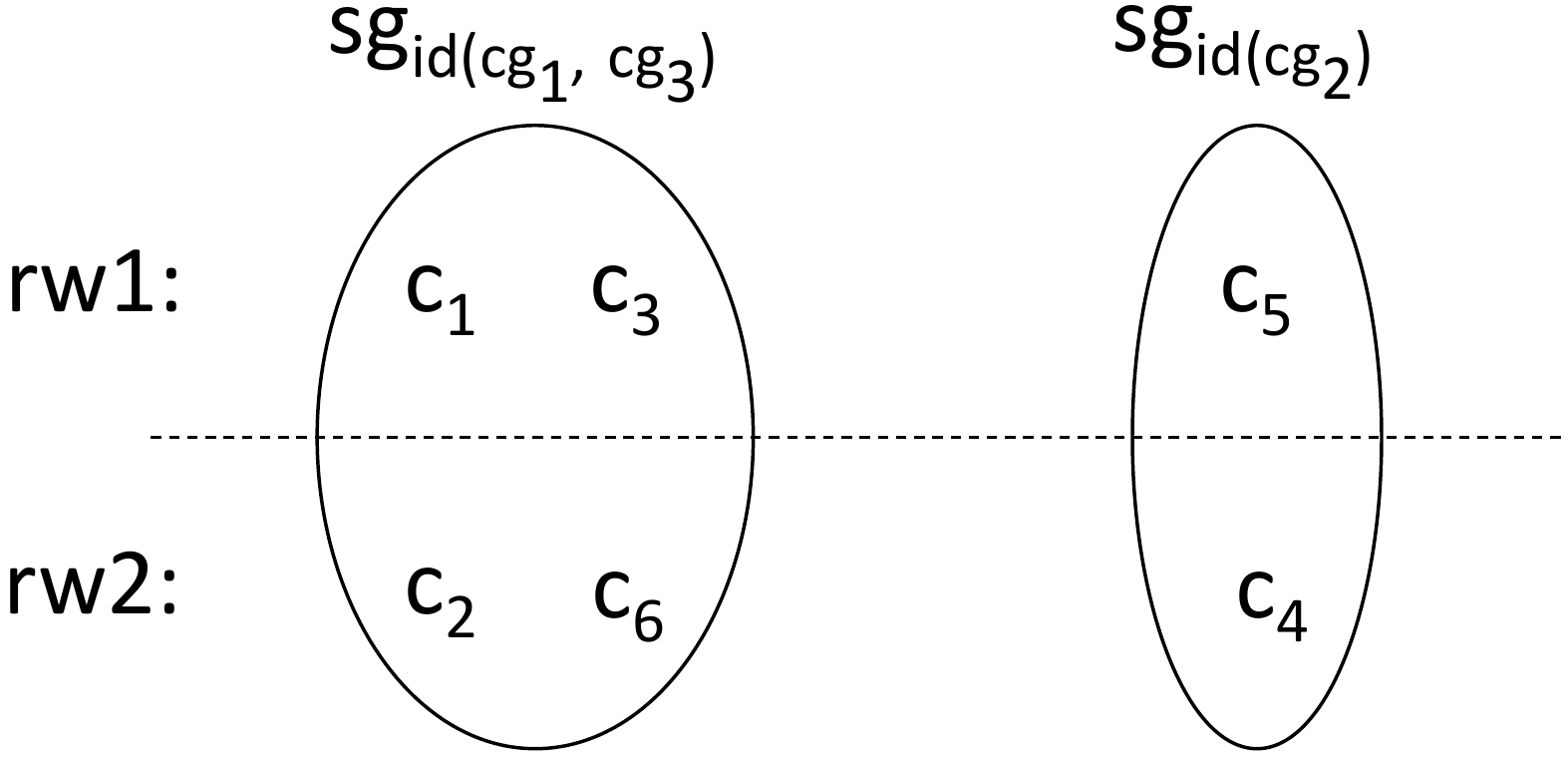}
\label{fig:subgraph4}}
\caption{Violation graph build example}
\label{fig:subgraph}
\end{figure}

An illustrative example is in order. Let's assume there are two RWs, $rw1$ and $rw2$, and the current violation graph consists in two subgraphs $sg_{id(cg_1)}$, containing cells $c_1, c_2, c_3$, and $sg_{id(cg_2)}$, containing cells $c_4, c_5$. 
In our example, the violation graph is partitioned as in Figure~\ref{fig:subgraph1}: both RWs have a portion of cells of every subgraph.

\subsubsection{The Coordinator}

The problem we address now is due to the dynamics of the violation graph, which evolves as new violation messages stream into the Repair module.
As each subgraph is partitioned among all RWs, subgraph partitions must be identified by the same ID: this is because a subgraph is a proxy for an equivalence class, and all its cells contribute to the correct functioning of the repair algorithm. 

Continuing with the example from Figure~\ref{fig:subgraph1}, suppose a new violation message $\{id(cg_3), c_6, c_1\}$ is received by all RWs. 
Now, in $rw1$, the new violation is added to subgraph $sg_{id(cg_1)}$ since both the message and the subgraph share the same cell $c_1$: as such, the new subgraph becomes $sg_{id(cg_1,cg_3)}$.
Instead, in $rw2$, the new violation triggers the creation of a new subgraph $sg_{id(cg_3)}$, since no common cells are shared between the message and existing subgraphs in $rw2$.
The violation graph becomes \emph{inconsistent}, as shown in Figure~\ref{fig:subgraph3}: this is a consequence of the independent operation of RWs. Instead, the repair algorithm requires the violation graph to be in a consistent state, as shown in Figure~\ref{fig:subgraph4}, where both RWs use the same subgraph identifier for the same equivalence class.

To guarantee the consistency of the violation graph among independent RWs,
Bleach uses a stateless coordinator component
that helps RWs agree on subgraph identifiers. In what follows we present three variants of the simple protocol RWs use to communicate with the coordinator.

\noindent \textbf{RW-basic.} When a RW receives violation messages for a tuple, 
it adds the cells in the messages to the violation graph, according to its local state.
Then, the RW creates a merge proposal containing the subgraph id
for each conflicting attribute, and sends it to the coordinator.  

Once the coordinator receives merge proposals from all RWs, it produces a merge decision, which contains a list of all cg IDs contained in the various merge proposals, and broadcasts it to all RWs. 
RWs merge their local subgraphs and converge to a globally consistent state.

Clearly, such a simple approach to coordination harms Bleach performance. Indeed, the RW-basic scheme requires one round-trip message for every incoming data tuple, from all RWs.

However, we note that it is not necessarily true that the coordination is always needed for every tuple.
For example, when every cell violates at most one rule, every subgraph would only have a single cg ID. Thus, coordination is not necessary.
More generally, given violation messages for a tuple, coordination is only necessary when there is a complete violation message containing an old cell which already exists in the violation graph because of a different violation rule.

\begin{figure}
	\centering
	\subfigure[initial state]{\includegraphics[width=0.45\linewidth]{subgraph1.pdf}\label{fig:subgraphexample21}}
	\quad
	\subfigure[after independent processing]{\includegraphics[width=0.45\linewidth]{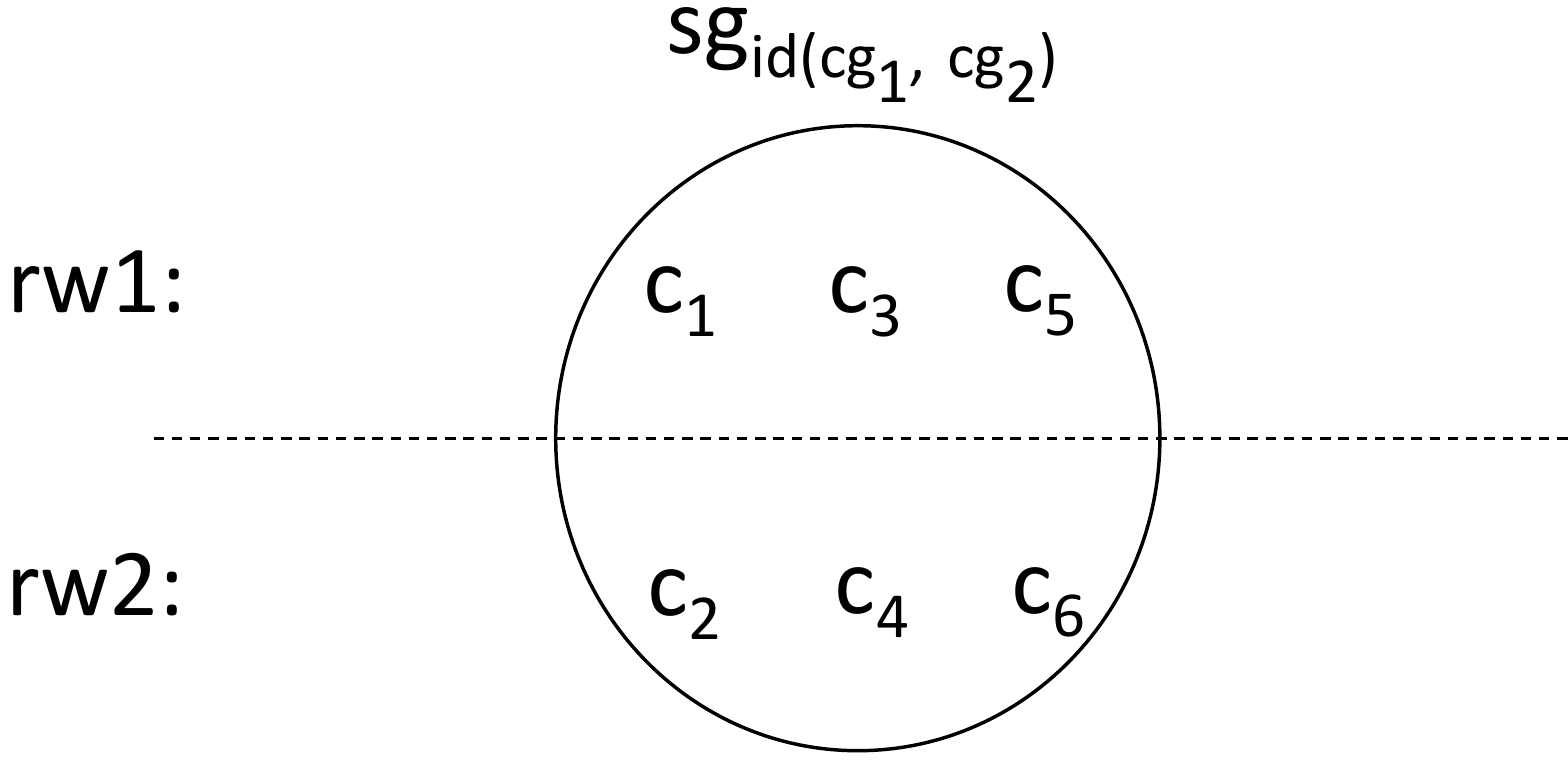}\label{fig:subgraphexample22}}
	\caption{Example of violation graph built without coordination}
	\label{fig:subgraphexample2}
\end{figure}

Figure~\ref{fig:subgraphexample2} gives an example, where the initial state (Figure~\ref{fig:subgraphexample21}) is the same as in Figure~\ref{fig:subgraph1}. 
Then, two violation messages, $\{id(cg_1), c_6, null\}$ and $\{id(cg_2), c_6, null\}$, are received.
Cell $c_6$ is a current cell contained in the current tuple. 
Obviously $sg_{id(cg_1)}$ and $sg_{id(cg_2)}$ should merge into $sg_{id(cg_1, cg_2)}$. 
This can be accomplished without coordination by both repair workers, as shown in Figure~\ref{fig:subgraphexample22}.
Indeed, each RW is aware that $c_6$ is involved in two subgraphs, although $c_6$ is only stored in $rw2$ because of the partitioning scheme.

Next, we use the above observations and propose two variants of the coordination mechanism that aim at bypassing the coordinator component to improve performance. In both variants, if there exist subgraphs which can merge correctly without coordination like the example of Figure~\ref{fig:subgraphexample2}, they merge immediately.

\noindent \textbf{RW-dr.} 
In RW-dr, the coordination is only conducted if it is necessary, and the repair worker sends a merge proposal to the coordinator and waits for the merge decision. 
However, this approach is not exempt from drawbacks: it may cause some data tuples in the stream to be delivered out of order.
This is because the repair worker wait for the merge decision in a non-blocking way.
The violation messages of a tuple which do not require coordination may be processed in the coordination gap of an earlier tuple.

\noindent \textbf{RW-ir.} 
With this variant, no matter if the violation messages of a tuple require coordination or not, a RW immediately updates its local subgraphs, executes the repair algorithm and emits a repair proposal downstream to the aggregator component. Then, if necessary, the RW lazily executes the coordination protocol.

Clearly, this approach caters to system performance and avoids tuples to be delivered out of order, but might harm cleaning accuracy. Indeed, individual data repair proposals from a RW are based on a local view 
prior to finishing all necessary merge operations on subgraphs, which has a direct impact on equivalence classes.

\subsubsection{The Aggregator}
Using the distributed violation graph, each RW executes independently the Bleach repair algorithm and emits a data repair proposal, which includes all\footnote{In case there are too many candidate values, we only send the top-$k$ values, where $k=5$.} candidate values and their frequency computed in a local subgraph partition.
The aggregator component collects all repair proposals and selects the candidate value to repair a given cell as the one having the highest aggregate frequency. Finally, the aggregator modifies the current data tuple and outputs a clean data stream.

Note that the aggregator only modifies current tuples. Instead, more importantly, the cells stored in the violation graph are not modified regardless of the repair decision: this allows to update frequency counts as new data streams into the system, thus steering the aggregator to make different repair decisions as the violation graph evolves.

%% file: 04-dynamicrule.tex
\section{Dynamic rule management}
\label{sec:dyn}

In stream data cleaning the rule set is usually not immutable but dynamic.
Therefore, we now introduce a new component, the rule controller, shown in Figure~\ref{fig:overview}, which allows Bleach to adapt to rule dynamics.
The rule controller accepts rule updates as input and guides the detect and the repair module to adapt to rule dynamics without stopping the cleaning process and without loosing state. Rule updates can be of two types: one for adding a new rule and one for deleting an existing rule.

\noindent \textbf{Detect.} In the detect module, the addition of a rule triggers the instantiation of a new DW, as input tuples are partitioned by rule. The new DW starts with no state, which is built upon receiving new input tuples. As such, violation detection using past tuples cannot be achieved, which is consistent with the Bleach design goals. Instead, the deletion of an existing rule simply triggers the removal of a DW, with its own local data history.

\noindent \textbf{Repair.} In the Repair module, the addition of a new rule is not problematic with respect to violation graph maintenance operations. Instead, the removal of a rule implies violation graph dynamics (subgraphs might shrink or split) which are more challenging to address.

Thus, in a subgraph, we further group cells by cell groups. 
A subgraph now can also be expressed as:

\noindent $sg_k=(id(cg_1,cg_2,...),[cg_1,cg_2,...])$, where
each cell group gathers super cells.
Some cells might span multiple groups, as they may violate multiple rules.
We label such peculiar cells as \emph{hinge cells}.
For each hinge cell, the subgraph keeps the IDs of its connecting cell groups:
$c^{*}_{i,j} = (c_{i,j}, id(cg_{i_1},cg_{i_2},...))$.
Hinge cells with the same value and the same connecting cell groups are also compressed into super cells.

With the new organization of cells in subgraphs, the violation graph updates as following upon the removal of a rule.
If a subgraph contains a single cell group related to the deleted rule, RWs are simply instructed to remove it.
If a subgraph contains multiple cell groups, RWs remove the cell groups related to the deleted rule and update the hinge cells.
With the remaining hinge cells, RWs check the connectivity of the remaining cell groups in the subgraph and decide to split the subgraph or not\footnote{A detailed algorithm can be found in our technical report: \url{http://www.eurecom.fr/~tian/bleach/bleachTR.pdf}}.

\begin{figure}[]
\centering
\subfigure[initial $sg_{id(cg_1, cg_2, cg_3)}$]{%
\includegraphics[width=0.45\linewidth]{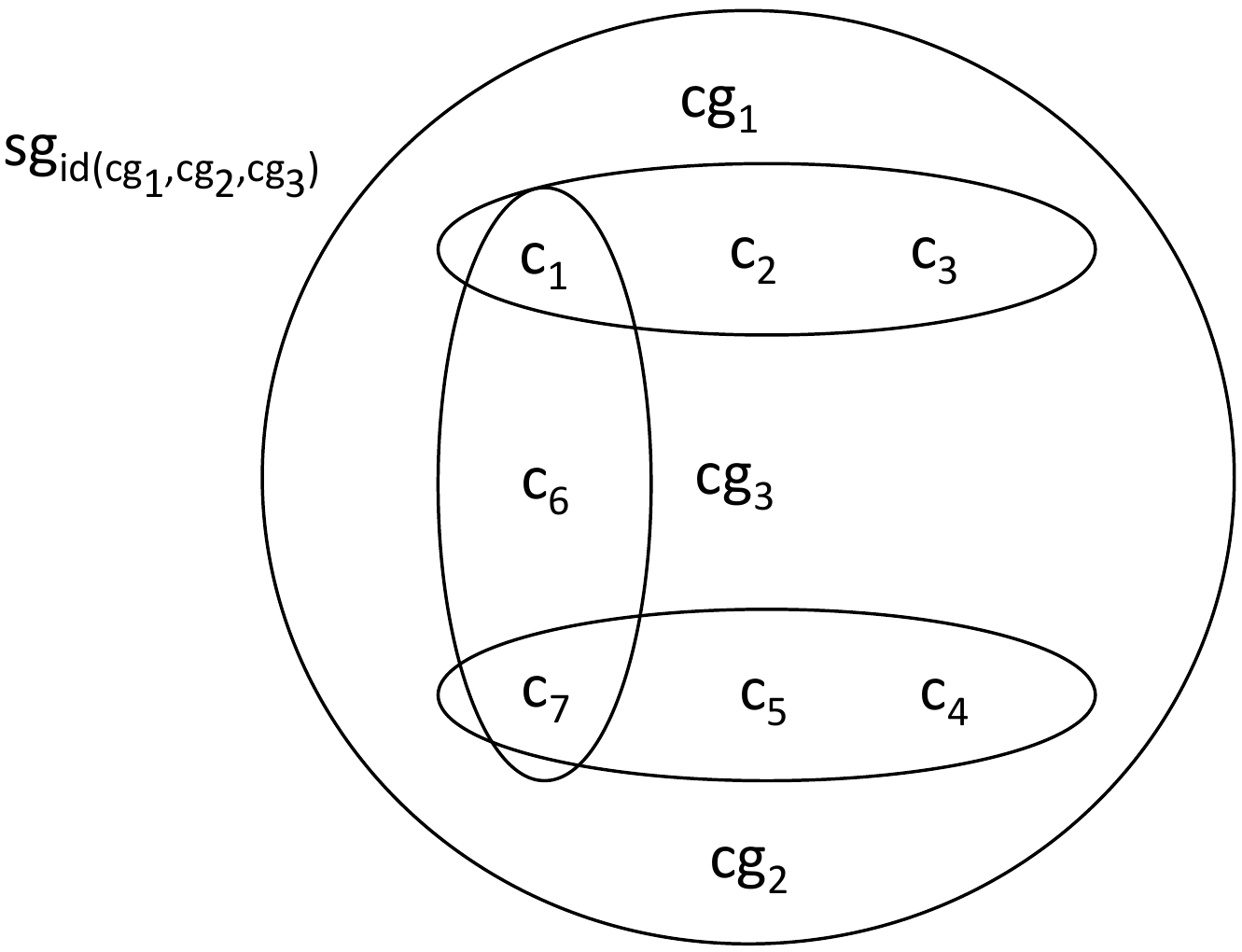}
\label{fig:drex1}}
\quad
\subfigure[if delete $cg_2$]{%
\includegraphics[width=0.45\linewidth]{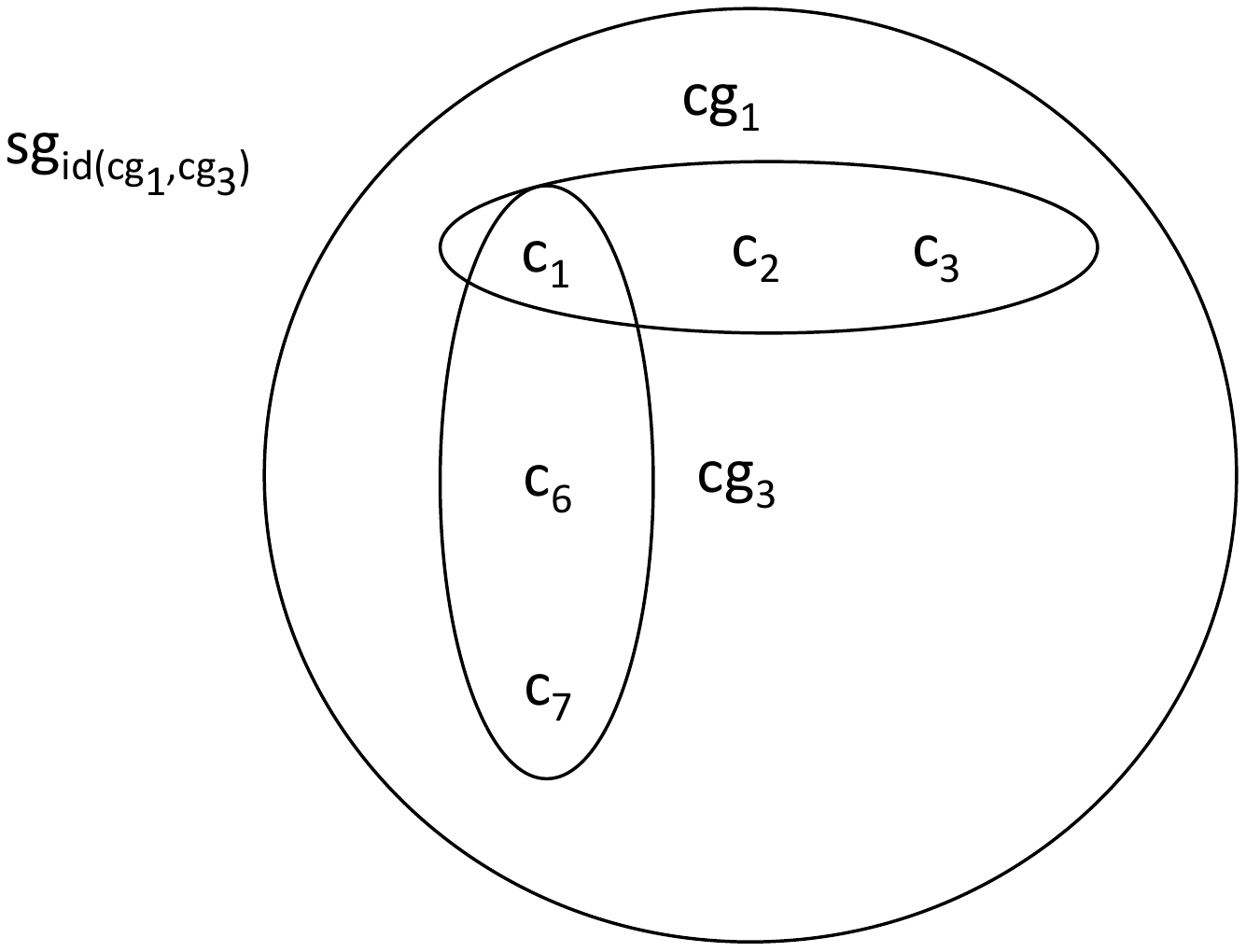}
\label{fig:drex2}}
\subfigure[incorrect if delete $cg_3$]{%
\includegraphics[width=0.45\linewidth]{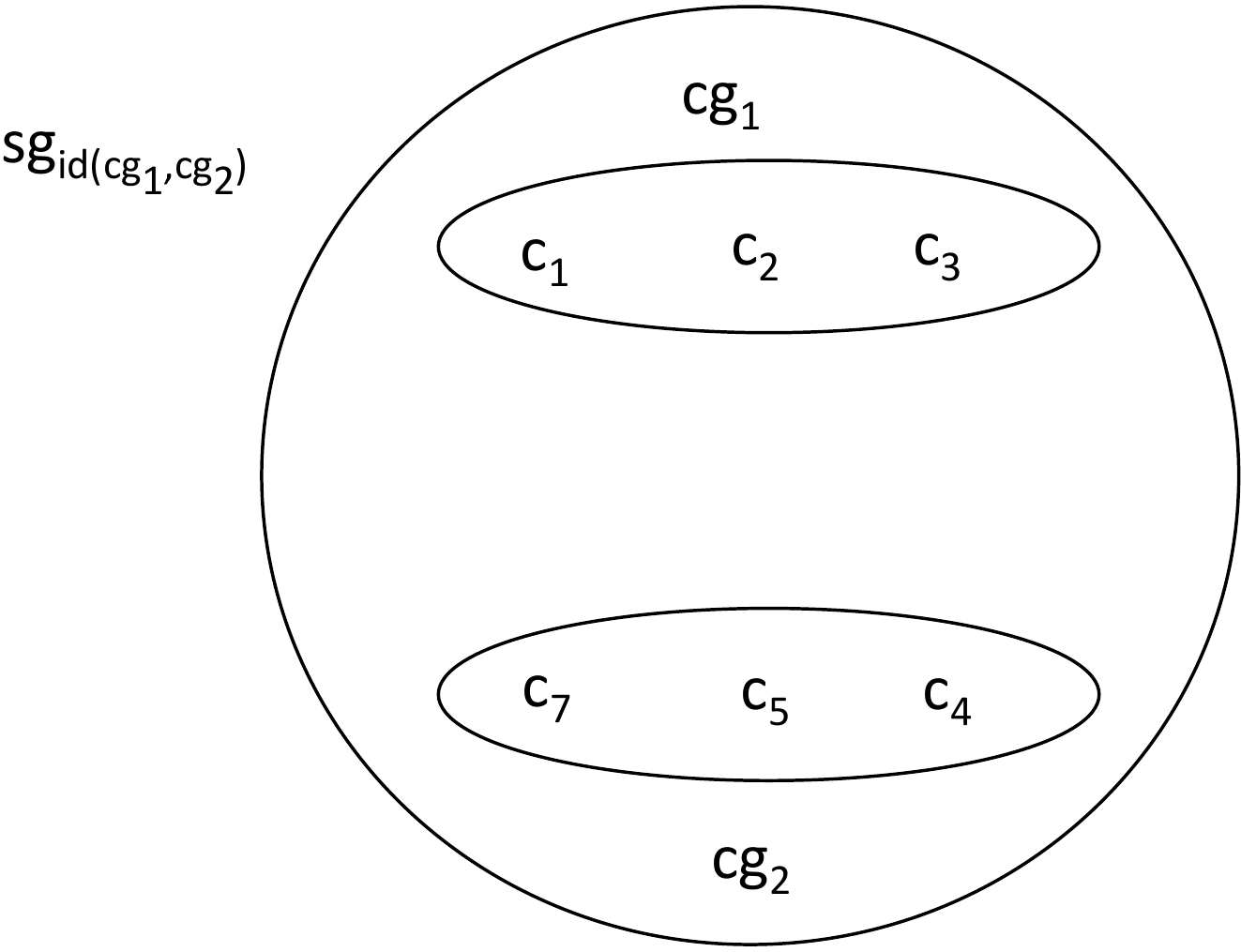}
\label{fig:drex3}}
\quad
\subfigure[correct if delete $cg_3$]{%
\includegraphics[width=0.45\linewidth]{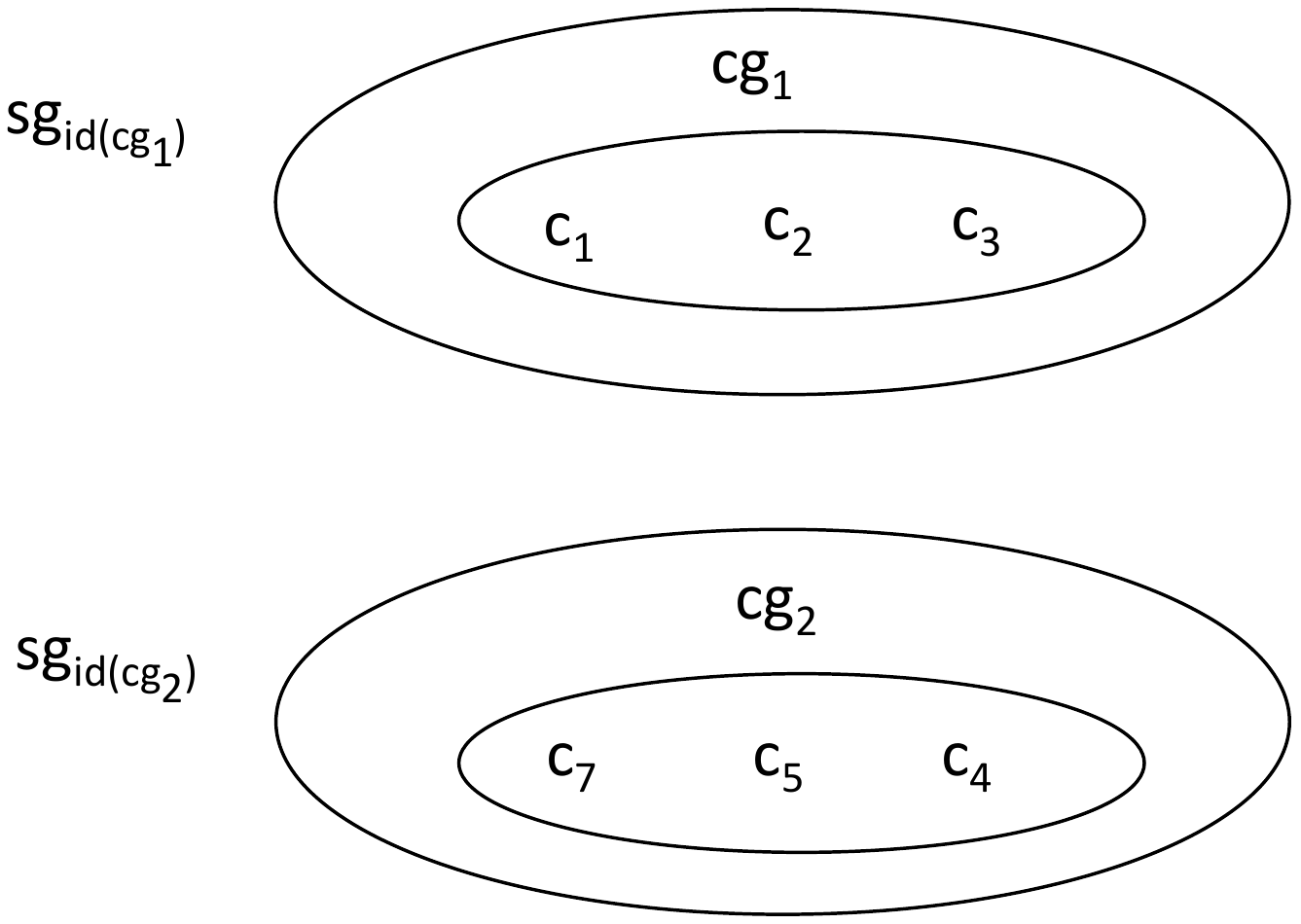}
\label{fig:drex4}}
\caption{Subgraph split example}
\label{fig:dynamicruleex}
\end{figure}

An example of a split operation can be seen in Figure~\ref{fig:dynamicruleex}. The initial state of a subgraph is shown in Figure~\ref{fig:drex1}: the subgraph is $sg_{id(cg_1, cg_2, cg_3)}$, and its contents are three cell groups. Cell $c_1$ and $c_7$ are \emph{hinge} cells, which work as bridges, connecting different cell groups together. 
Now, as a simple case, assume we want to remove the rule pertaining to $cg_2$: the subgraph should become $sg_{id(cg_1,cg_3)}$, as shown in Figure~\ref{fig:drex2}. Note that cell $c_7$ looses its status of hinge cell.
A more involved case arise when we delete the rule pertaining to $cg_3$ instead of the rule pertaining to $cg_2$. In this case, the subgraph should not become $sg_{id(cg_1,cg_2)}$ as shown in Figure~\ref{fig:drex3}. Indeed, removing $cg_2$ eliminates all existing hinge cells connecting the remaining cell groups. Thus, the subgraph must split in two separate subgraphs $sg_{id(cg_1)}$ and $sg_{id(cg_2)}$ as shown in Figure~\ref{fig:drex4}.

%% file: 05-windowing.tex
\section{windowing}
\label{sec:window}
Bleach provides windowed computations, which allow expressing data cleaning over a sliding window of data. Despite being a common operation in most streaming systems, window-based data cleaning addresses the challenge of the unbounded nature of streaming data: without windowing, the data structures Bleach uses to detect and repair a dirty stream would grow indefinitely, which is unpractical.

In this section, we discuss two windowing strategies: a basic, tuple-based windowing strategy and an advanced strategy that aim at improving cleaning accuracy.

\subsection{Basic Windowing}
The underlying idea of the basic windowing strategy is to only use tuples within the sliding window to populate the data structures used by Bleach to achieve its tasks. Next, we outline the basic windowing strategy for both DWs and RWs operation.

\noindent \textbf{Windowed Detection.} We now focus on how DWs maintain their local data history. Clearly, the data history only contains cells that fall within the current window. When the window slides forward, DWs update the data history as follows: {\it i)} if a cell group ends up having no cells in the new window, DWs simply delete it; {\it ii)} for the remaining cell groups, DWs drop all cells that fall outside the new window, and update accordingly the remaining super cells.

Note that, if implemented naively, the first operation above can be costly as it involves a linear scan of all cell groups. To improve the efficiency of data history updates, Bleach uses the following approach. It creates a FIFO queue of $k$ lists, which store cell groups. 
In case the sliding step is half the window size, $k=2$; more generally, we set $k$ to be the window size divided by the sliding step.
Any new cell group from the current window enters the queue in the $k$-th list. Any cell group updates, e.g. due to a new cell added to the cell group, ``promotes'' it from list $j$ to list $k$. As the window slides forward, the queue drops the list (and its cell groups) in the first position and acquires a new empty list in position $k+1$. 

\noindent \textbf{Windowed Repair.} Now we focus on how to maintain the violation graph in RWs. Again, the violation graph only stores cells within the current window. When the window slides forward, RWs update the violation graph as follows: 
\begin{itemize}
	\item If a subgraph has no cells in the new window, RWs delete the subgraph;
	\item For the remaining subgraphs, if a cell group has no cells in the new window, RWs delete the cell group;
	\item RWs also delete hinge cells that are outside of the new window. This could require subgraphs to split, as they could miss a ``bridge'' cell to connect its cell groups;
	\item For the remaining cell groups, RWs drop all cells outside of the new window, and update the remaining super cells accordingly.
\end{itemize}
For efficiency reasons, Bleach uses the same $k$-list approach described for DWs to manage violation graph updates due to a sliding window.

\begin{figure}
	\centering
	\subfigure[input data]{\includegraphics[width=0.63\linewidth]{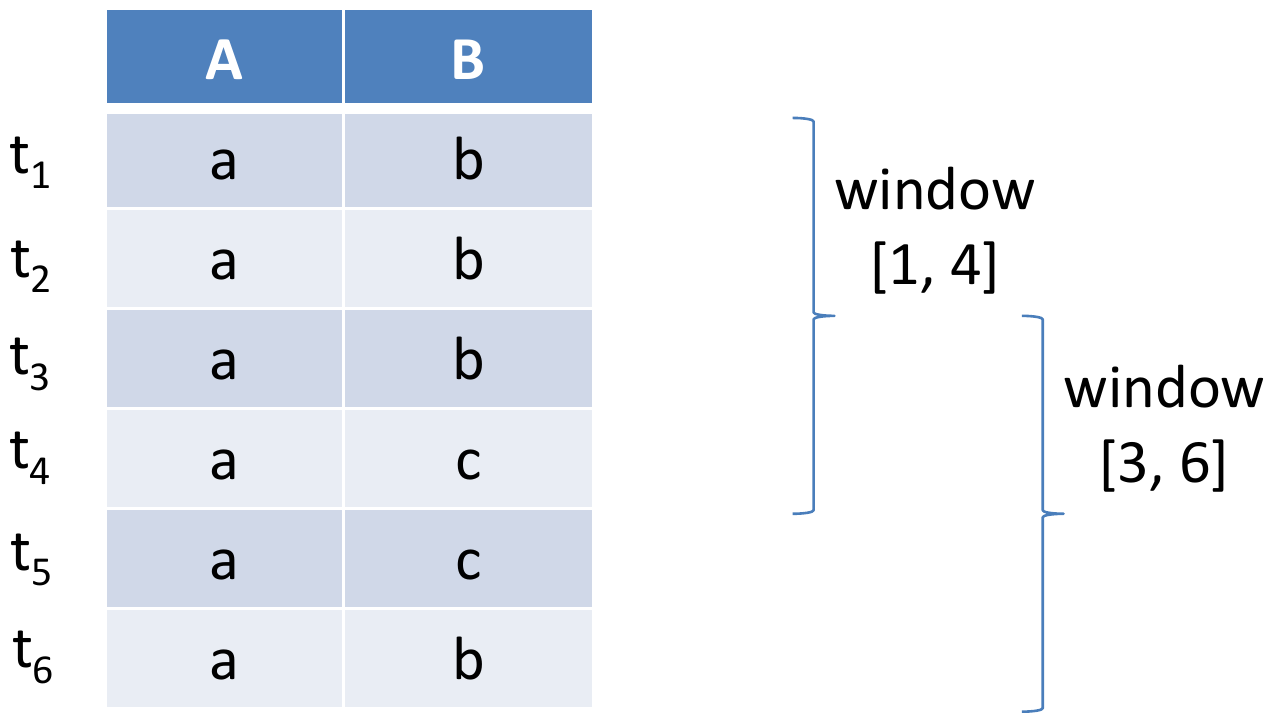}\label{fig:bwexample1}}
	\subfigure[output data with basic windowing]{\includegraphics[width=0.47\linewidth]{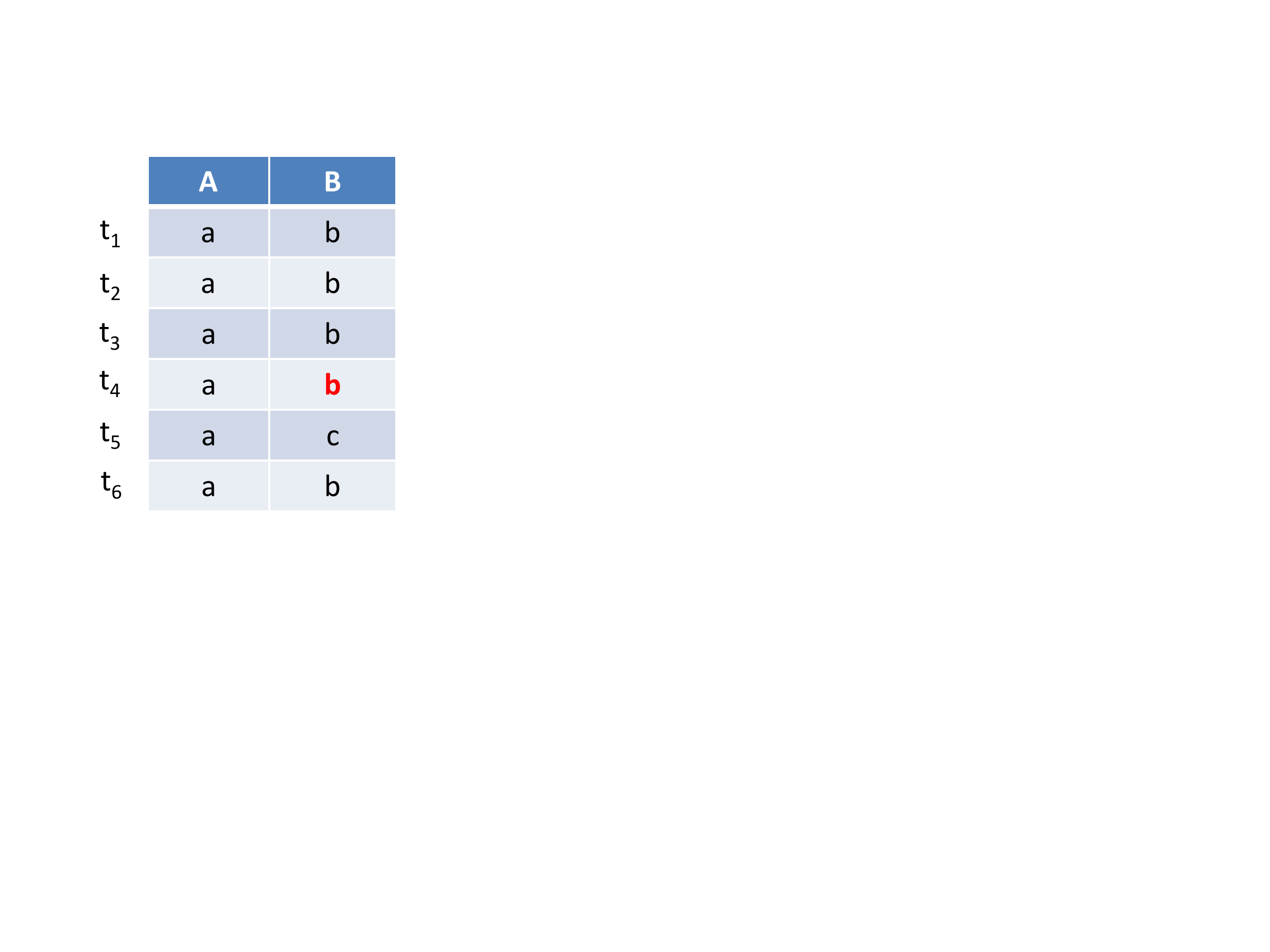}\label{fig:bwexample2}}
	\subfigure[output data with Bleach windowing]{\includegraphics[width=0.47\linewidth]{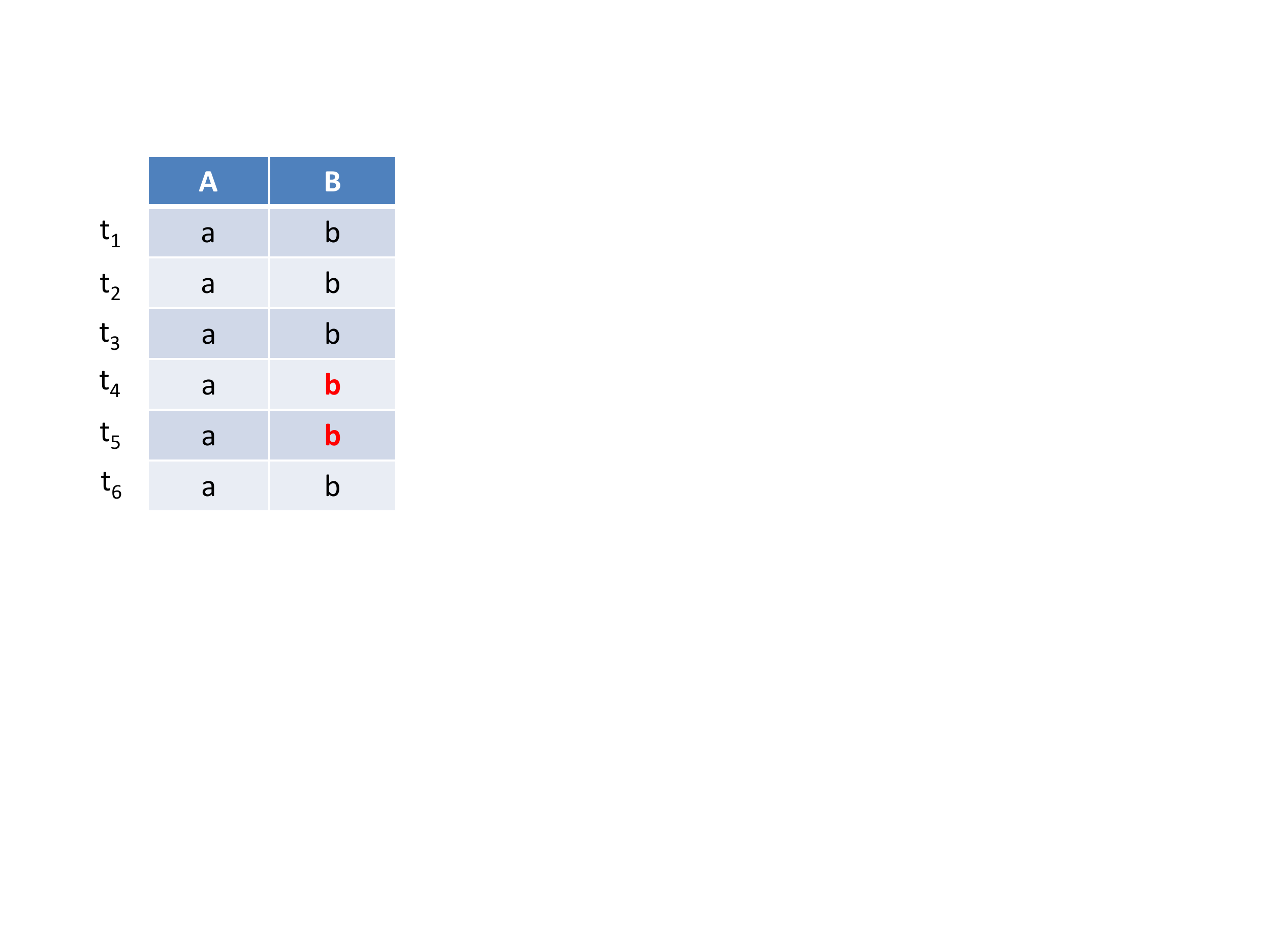}\label{fig:bwexample3}}
	\caption{Motivating example: Basic vs. Bleach windowing.}
	\label{fig:bwexample}
\end{figure}

\subsection{Bleach Windowing}
The basic windowing strategy only relies on the data within the current window to perform data cleaning, which may limit the cleaning accuracy. We begin with a motivating example, then describe the Bleach windowing strategy, that aims at improving cleaning accuracy. Note that here we only focus on the repair module and its violation graph, since Bleach windowing does not modify the operation of the detect module.

Figure~\ref{fig:bwexample1} illustrates a data stream of two-attribute tuples. Assume we use a single FD rule $(A \rightarrow B)$, a window size of 4 tuples, a sliding step of 2 tuples, and the \emph{basic} windowing strategy.

When $t_4$ arrives, the window covers tuples $[1,4]$. According to the repair algorithm, Bleach repairs $t_4(B)$ and sets it to the value $b$.
Now, when tuple $t_5$ arrives, the window moves to cover tuples $[3,6]$, even though $t_6$ has yet to arrive. With only three tuples in the current window,  the algorithm determines $t_5(B)$ is correct, because now the majority of tuples have value $c$. The output stream produced using basic windowing is shown in Figure~\ref{fig:bwexample2}. Clearly, cleaning accuracy is sacrificed, since it is easy to see that $t_5(B)$ should have been repaired to value $b$, which is the most frequent value overall. Hence, the need for a different windowing strategy to overcome such problems.
 
Bleach windowing relies on an extension of a super cell, which we call a \emph{cumulative super cell}. The idea is for the violation graph to accumulate past state, to complement the view Bleach builds using tuples from the current window. Hence, a cumulative super cell is represented as a super cell, with an additional field that stores the number of occurrences of cells (both hinge as well as super cells within a cell group) with the same RHS value, including those that have been dropped because they fall outside the sliding window boundaries.

When using Bleach windowing, RWs maintain the violation graph by storing cumulative super cells. When the window slides forward, RWs update the violation graph as follows. The first two steps are equivalent to those for the basic strategy. The last two steps are modified as follows:
\begin{itemize}
	\item For the remaining subgraphs, RWs update hinge cells, and ``flush'' those that do not bridge cell groups anymore because of the update. Also, RWs split subgraphs according to the remaining hinge cells;
	\item For the remaining cell groups and hinge cells, RWs update compressed super cells, ``flushing'' cells which fall outside the new window while keeping their count.
\end{itemize}
The ``flush'' operation used above only drops the content of a super cell, but keeps its structure and its count field. 

Now, going back to the example in Figure~\ref{fig:bwexample1}, when tuple $t_5$ arrives, Bleach stores two cumulative super cells: $csc_1(id(t)=3,value=`b\textrm',count=3)$ and $csc_2(id(t)=[4,5],value=`c\textrm',count=2)$. Although $t_1$ and $t_2$ have been deleted because they are outside the sliding window, they still contribute to the count field in $csc_1$. Therefore, tuple $t_5(B)$ is correctly repaired to value $b$, as shown in Figure~\ref{fig:bwexample3}.

\noindent \textbf{Additional notes:} When using cumulative super cells, Bleach keeps tracks of candidate values to be used in the repair algorithm as long as cell groups remain.
By using cumulative super cells for hinge cells, subgraphs only split if some cell groups are removed when the window moves forward. Note that the introduction of cumulative super cells does not interfere with dynamic rule management: in particular, when deleting a rule, subgraphs update correctly when hinge cells use the cumulative format. Overall, to compute the count of a candidate value in a subgraph, cumulative super cells accumulate the counts of relevant compressed super cells from all cell groups, taking into account any duplicate contributions from hinge cells.

Obviously, Bleach windowing requires more storage than basic windowing, as cumulative super cells store additional information, and because of the ``flush'' operation described earlier, which keeps a super cell structure, even when it has an empty cell list. Section~\ref{sec:eva} demonstrates that such additional overhead is well balanced by superior cleaning accuracy, making Bleach windowing truly desirable.

%% file: 06-evaluation.tex
\section{Evaluation}
\label{sec:eva}
Bleach prototype implementation is built using Apache Storm~\cite{storm}.\footnote{Nothing prevents Bleach to be built using alternative systems such as Apache Flink, for example.} Input streams, including both the data stream and rule updates, are fed into Bleach using Apache Kafka~\cite{kafka}.

Our goal is to demonstrate that Bleach achieves efficient stream data cleaning 
under real-time constraints. Our evaluation uses {\it throughput, latency} and {\it dirty ratio} as performance metrics. We express the dirty ratio as the fraction of dirty data remaining in the output data stream: the smaller the dirty ratio, the higher the cleaning accuracy. The processing latency is measured from uniformed sampled tuples (1 per 100). All our experiments are conducted in a cluster of 18 machines, with 4 cores, 8 GB RAM and 1 Gbps network interface each.

We evaluate Bleach using a synthetic dataset generated from TPC-DS (with scale factor 100 GB). To do so, we join a fact table \emph{store\_sales} with its dimension tables in TPC-DS to build a single table (288 million tuples). By exporting this table to Kafka, we simulate an ``unbounded'' data stream. 
We manually design eight CFD rules, as shown in Table~\ref{table:ruleset}. Among these rules, $r_4$ and $r_5$ have the same RHS attribute $s\_store\_name$, while $r_6$ and $r_7$ have the same RHS attribute $c\_email\_addr$, as intersecting attributes.

\begin{table}
\small
\centering
\caption{Rule Sets}
\begin{tabular}{ l c r }
\hline
$r_0:ss\_item\_sk \rightarrow i\_brand,(ss\_item\_sk \neq null)$\\
$r_1:ss\_item\_sk \rightarrow i\_category,(ss\_item\_sk \neq null)$\\
$r_2:ca\_state, ca\_city \rightarrow ca\_zip, (ca\_state, ca\_city \neq null)$ \\
$r_3:ss\_promo\_sk \rightarrow p\_promo\_name,(ss\_promo\_sk \neq null)$\\
$r_4:ss\_store\_sk \rightarrow s\_store\_name,(ss\_store\_sk \neq null)$\\
$r_5:ss\_ticket\_num \rightarrow s\_store\_name,(ss\_ticket\_num \neq null)$\\
$r_6:ss\_ticket\_num \rightarrow c\_email\_addr,(ss\_ticket\_num \neq null)$ \\
$r_7:ss\_customer\_sk \rightarrow c\_email\_addr,(ss\_customer\_sk \neq null)$\\
  \hline
\end{tabular}
\label{table:ruleset}
\end{table}

We generate a dirty data stream, according to our rules, as follows: we modify the values of RHS attributes with probability 10\% and replace the values of LHS attributes with NULL with probability 10\%.\footnote{In our experiments we also use BART~\cite{arocena2015messing}, which is a well accepted dirty data generator. However, due to the sheer size of our data stream, we present results obtained using our custom process, which mimics that of BART but scales to large data streams.}
In all the experiments, we set the window size to 2 M and the sliding step to 1 M tuples respectively, regardless which windowing strategy we use. If not otherwise specified, we use Bleach windowing as the default strategy.

\subsection{Comparing Coordination Approaches}
\label{ssec:exp1}

\begin{figure*}
\centering
\subfigure[Throughput\label{fig:exp1_throughput}]{\includegraphics[scale=0.35]{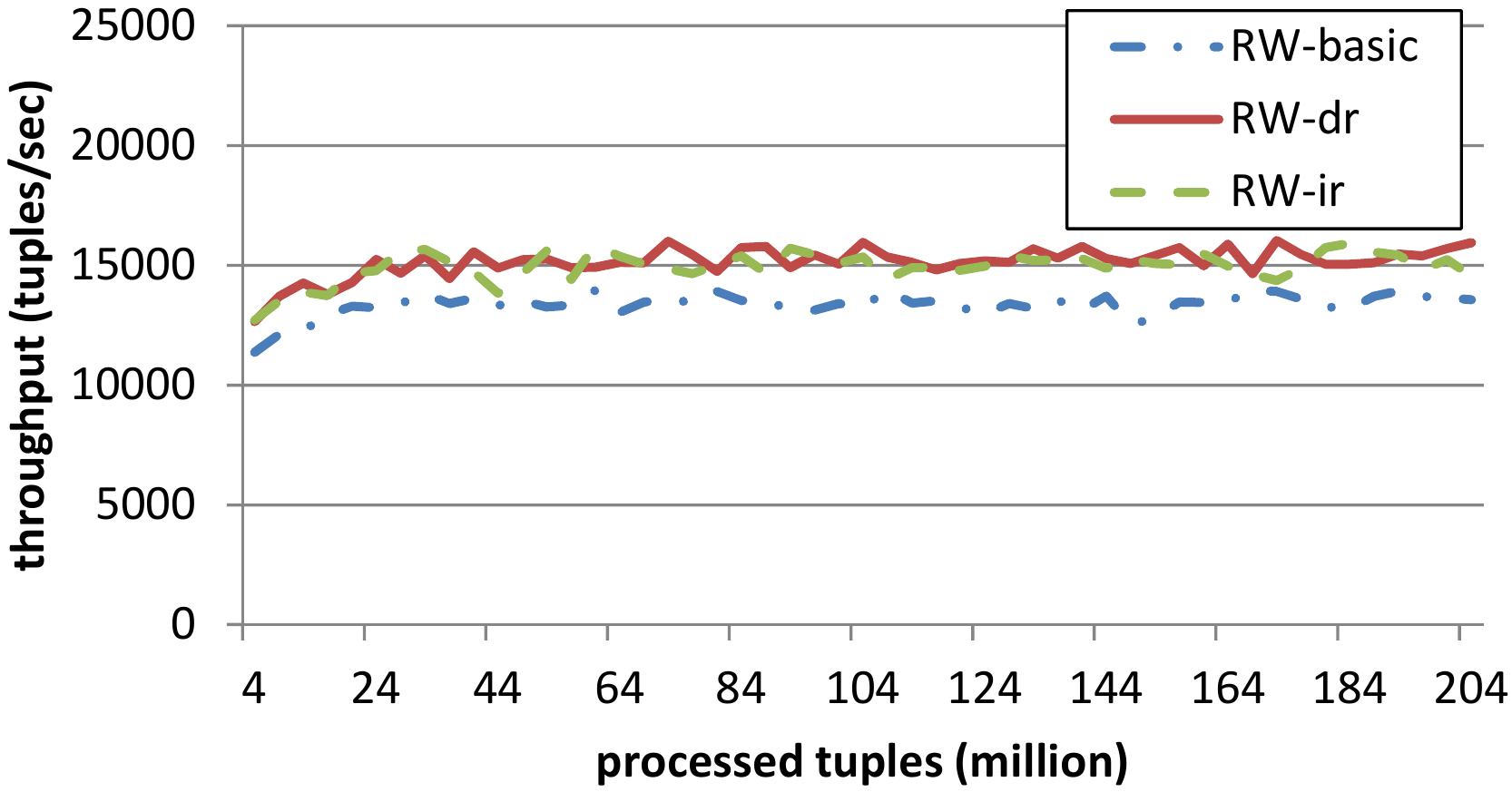}}
\subfigure[Latency\label{fig:exp1_latency}]{\includegraphics[scale=0.45]{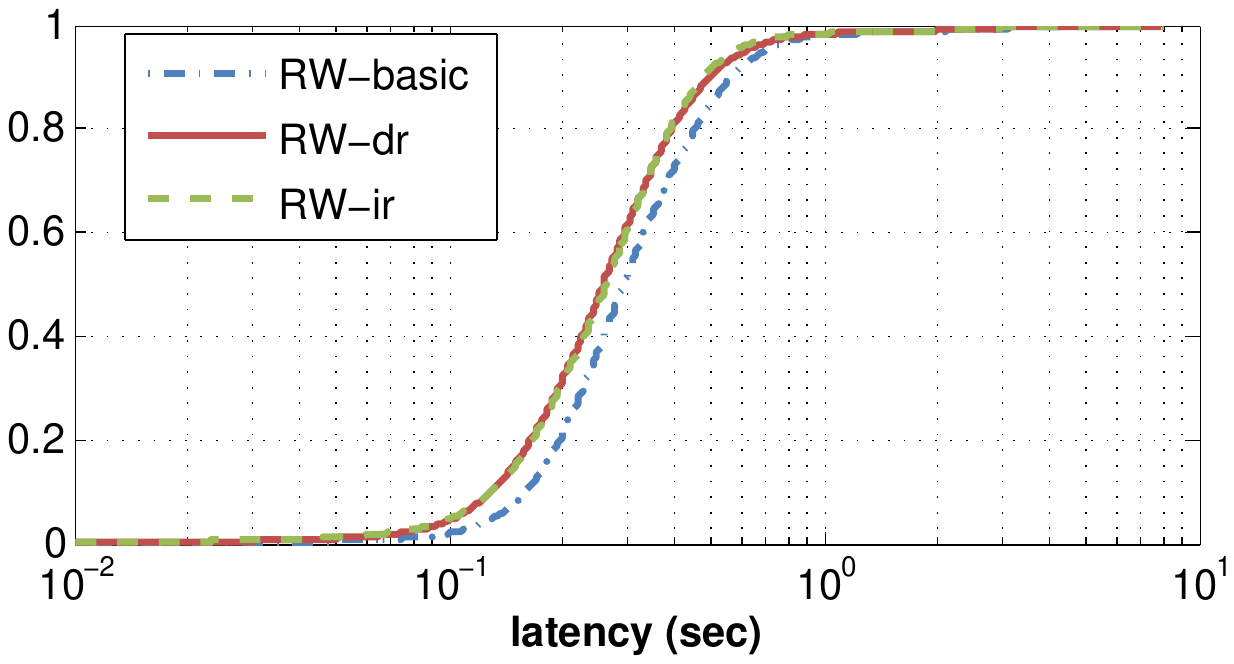}}
\subfigure[Dirty Ratio\label{fig:exp1_accuracy}]{\includegraphics[scale=0.35]{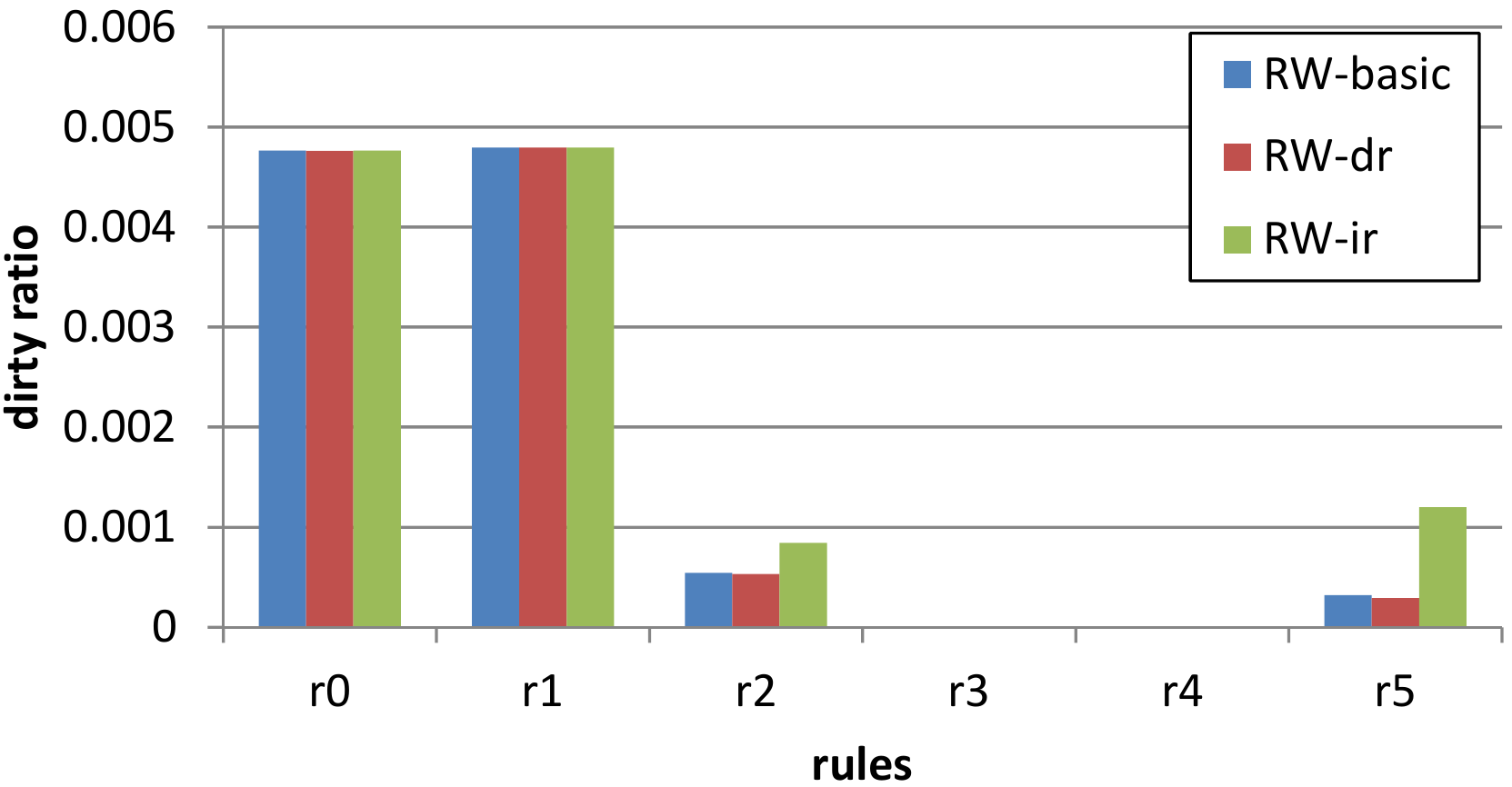}}
\caption{Comparison of coordination mechanisms: RW-basic, RW-dr and RW-ir.}
\label{fig:coordination_comparison}
\end{figure*}

In this experiment we compare the three RW approaches discussed in Section~\ref{ssec:repair}, according to our performance metrics, as shown in Figure~\ref{fig:coordination_comparison}: \emph{RW-basic} requires coordination among repair workers for each tuple; \emph{RW-dr} omits coordination for tuples when possible; \emph{RW-ir} is similar to \emph{RW-dr}, but allows repair decisions to be made before finishing coordination. Next, we use our synthetic dataset and rules $r_0$ to $r_5$.

Figure~\ref{fig:exp1_throughput} shows how Bleach throughput evolves with processed tuples. The throughput with both \emph{RW-dr} and \emph{RW-ir} is around 15K tuples/second, whereas \emph{RW-basic} achieves roughly 13K tuples/second. The inferior performance of RW-basic is due to the large number of coordination messages required to converge to global subgraph identifiers, while \emph{RW-dr} and \emph{RW-ir} only require 7\% coordination messages in \emph{RW-basic}.

Figure~\ref{fig:exp1_latency} shows the CDF of the tuple processing latency for the three RW approaches. \emph{RW-basic} has the highest processing latency, on average 364 ms. The processing latency of \emph{RW-ir} is on average 316 ms. \emph{RW-dr} average latency is slightly higher, about 323 ms.
This difference is due again to the additional round-trip-messages required by coordination: with \emph{RW-ir}, RWs make their repair proposals without waiting for coordination to complete, therefore the small processing latency. 

Figure~\ref{fig:exp1_accuracy} illustrates the cleaning accuracy. All three approaches lower the ratio of dirty data significantly, from the initial 10\% to at most 0.5\% (even 0\% for rule $r_3$ and $r_4$). 
For the first five rules, the three approaches achieve similar cleaning accuracy. Instead, for rule $r_5$ the \emph{RW-ir} method suffers and the dirty ratio is larger. 
Indeed, for rule $r_5$ whose cleaning accuracy is heavily linked to rule $r_1$, \emph{RW-ir} fails to correctly update some of its subgraphs because it eagerly emits repair proposals without waiting for coordination to complete.

\subsection{Comparing Windowing Strategies}

In this experiment, we compare the performance of the basic and Bleach windowing strategies, and use the \emph{RW-dr} mechanism. As for the previous experiment, we use rules $r0$ to $r5$. Additionally, for stress testing, we increase the input dirty data ratio from 10\% to 50\% for data in the interval from 40 M to 42 M tuples.

\begin{figure}
	\centering
	{\includegraphics[width=0.66\linewidth]{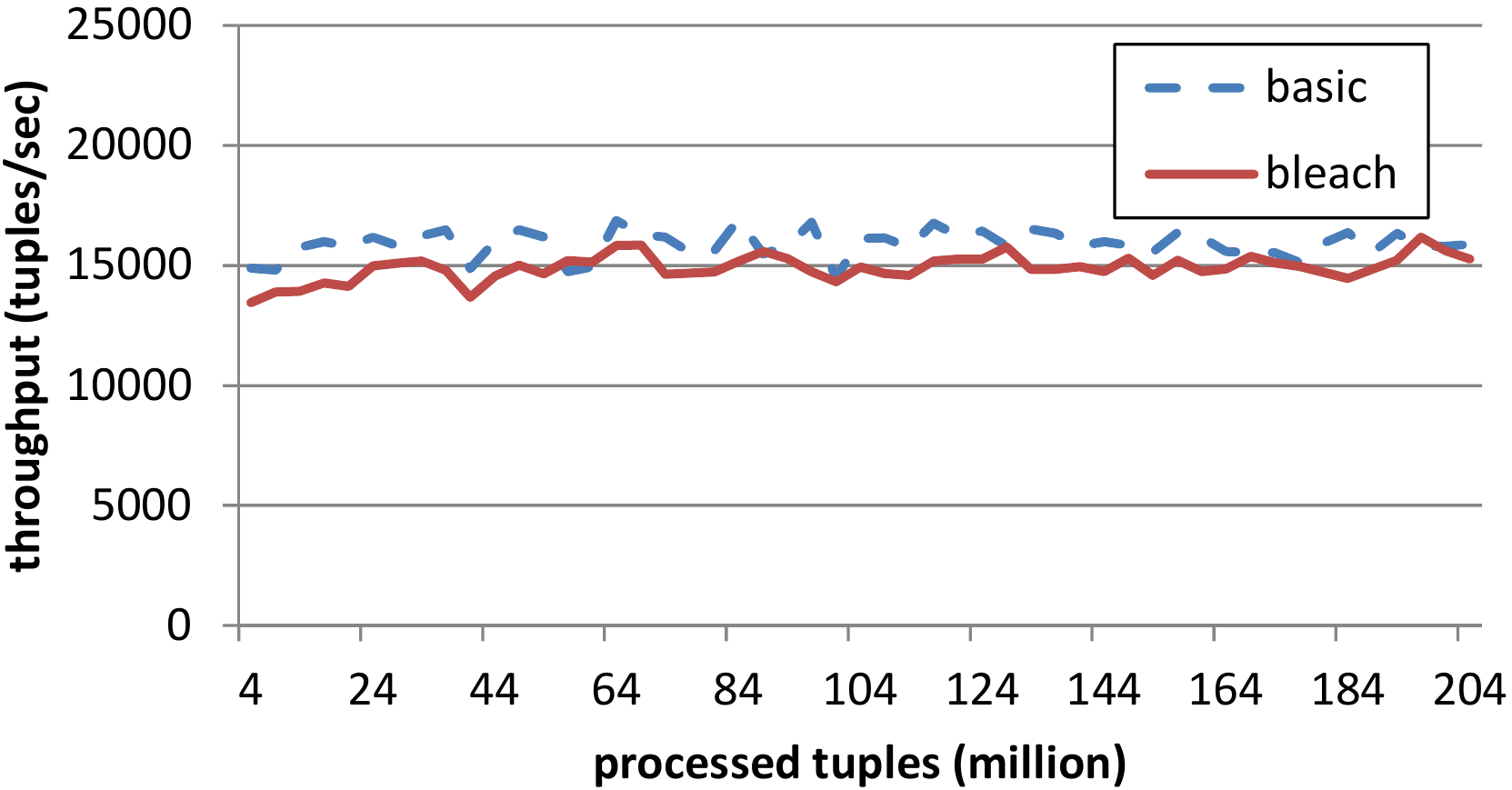}}
	\caption{throughput of two windowing strategies}
	\label{fig:exp2_throughput}
\end{figure}

\begin{figure}
	\centering
	{\includegraphics[width=0.66\linewidth]{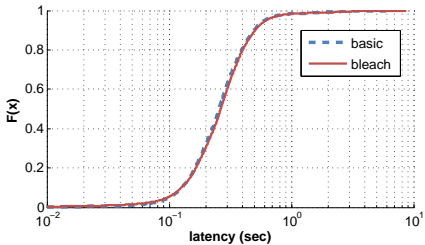}}
	\caption{processing latency CDF of two windowing strategies}
	\label{fig:exp2_latency}
\end{figure}

\begin{figure*}[t]
	\centering
	\subfigure[r0]{\includegraphics[width=0.32\linewidth]{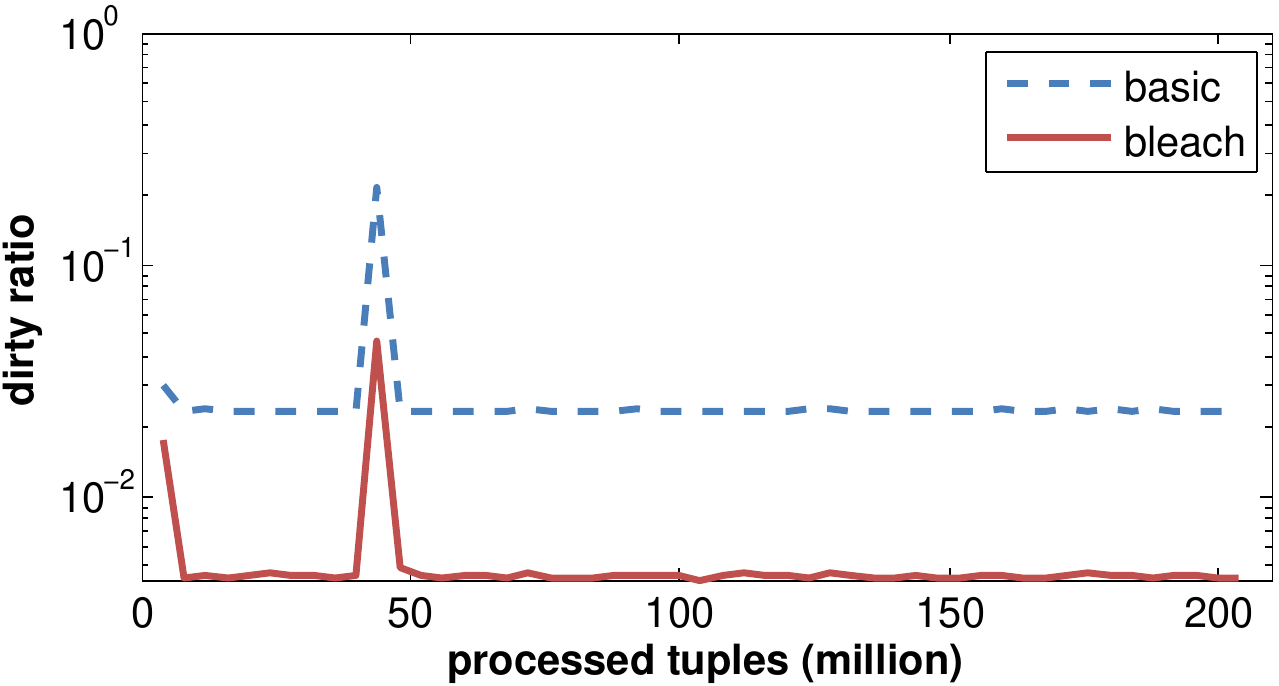}\label{fig:exp2_a_r0}}
	\subfigure[r1]{\includegraphics[width=0.32\linewidth]{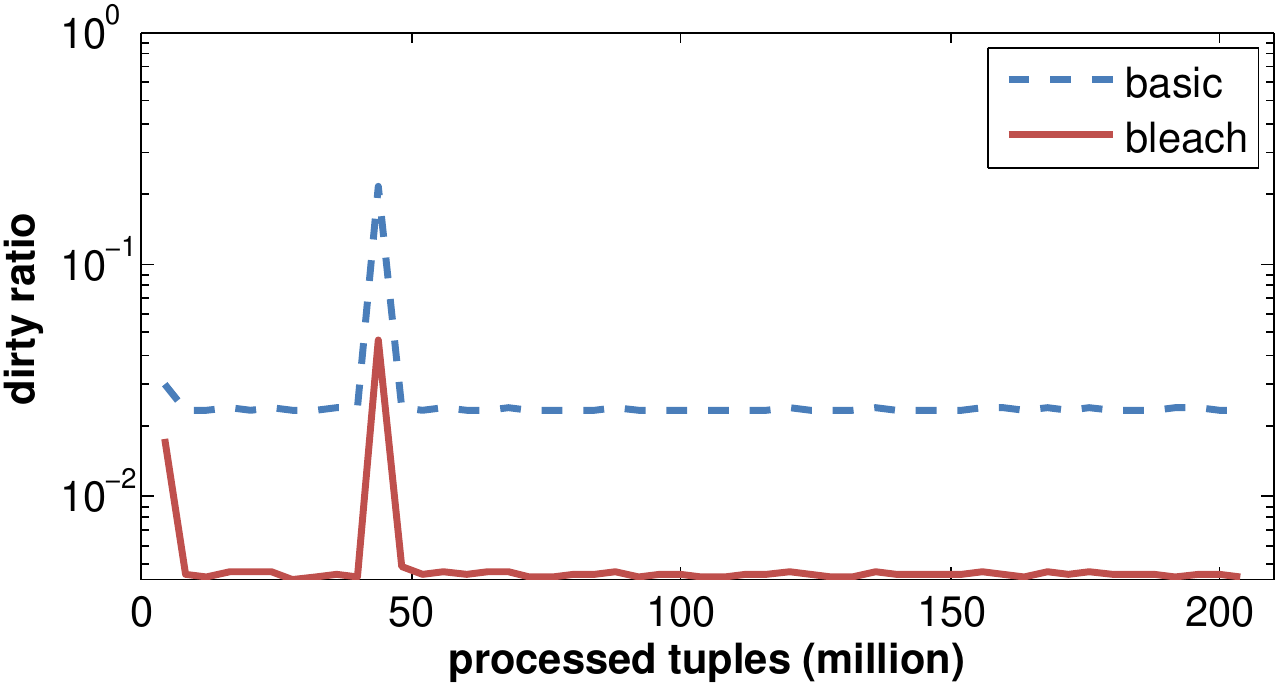}\label{fig:exp2_a_r1}}
	\subfigure[r2]{\includegraphics[width=0.32\linewidth]{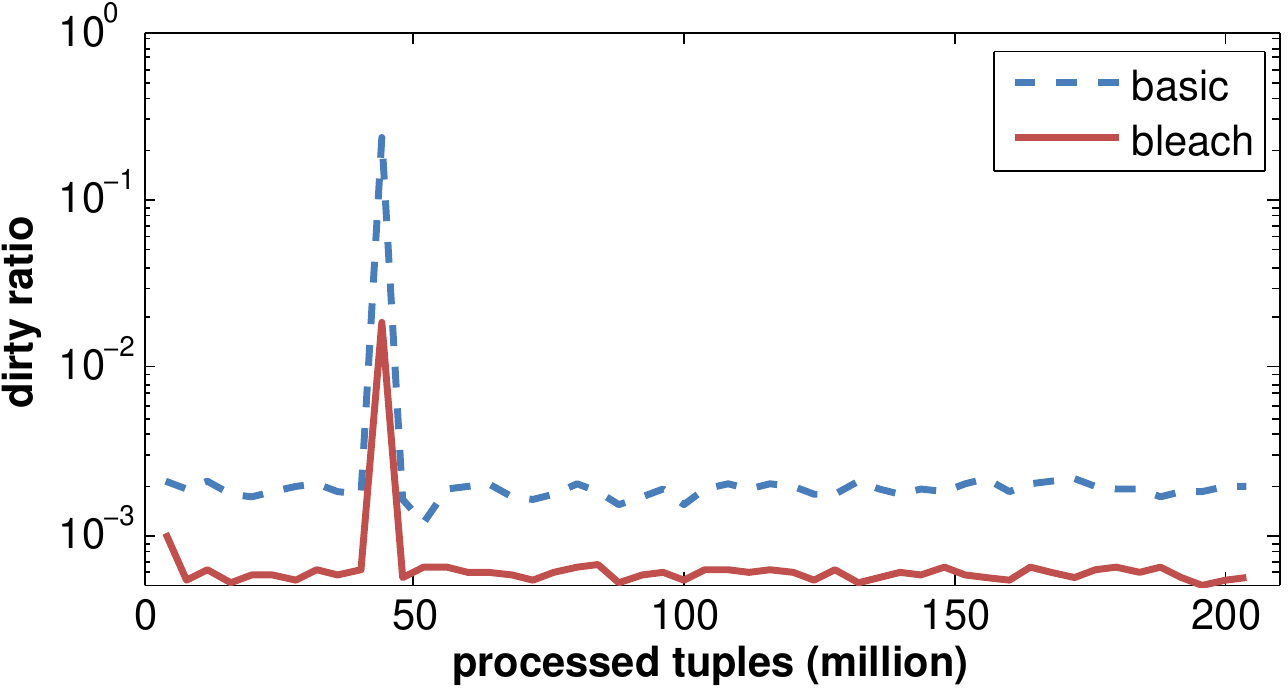}\label{fig:exp2_a_r2}}
	\subfigure[r3]{\includegraphics[width=0.32\linewidth]{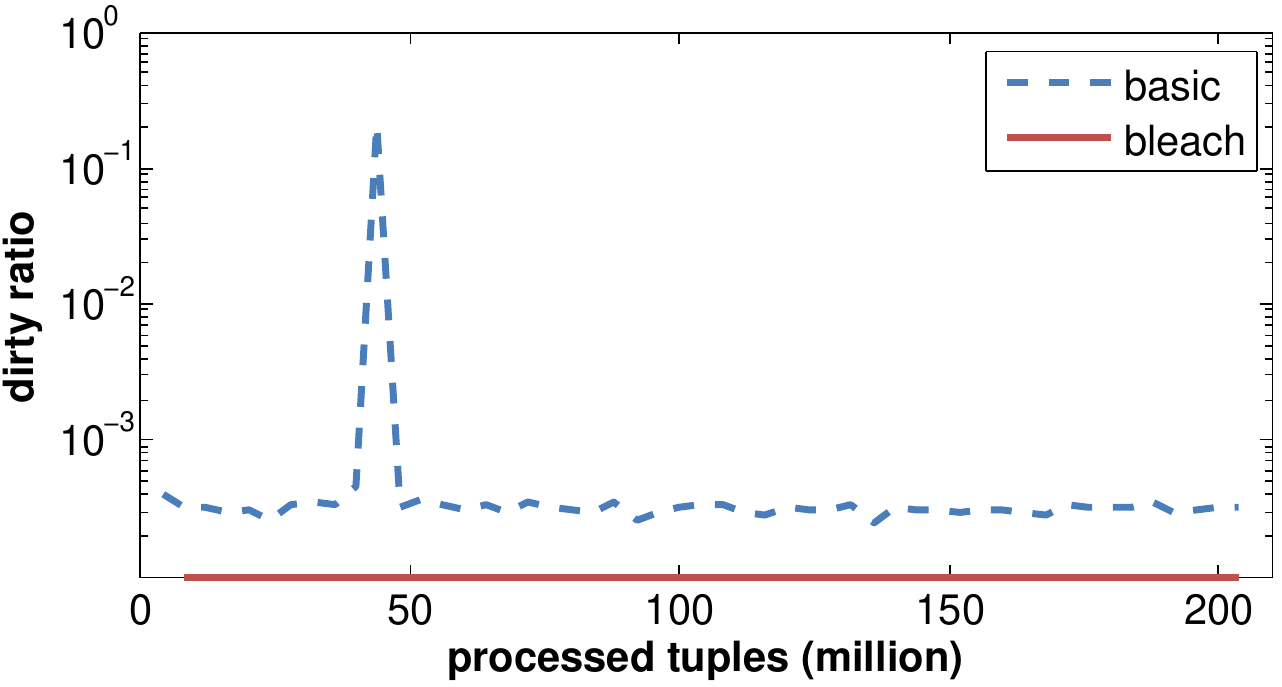}\label{fig:exp2_a_r3}}
	\subfigure[r4]{\includegraphics[width=0.32\linewidth]{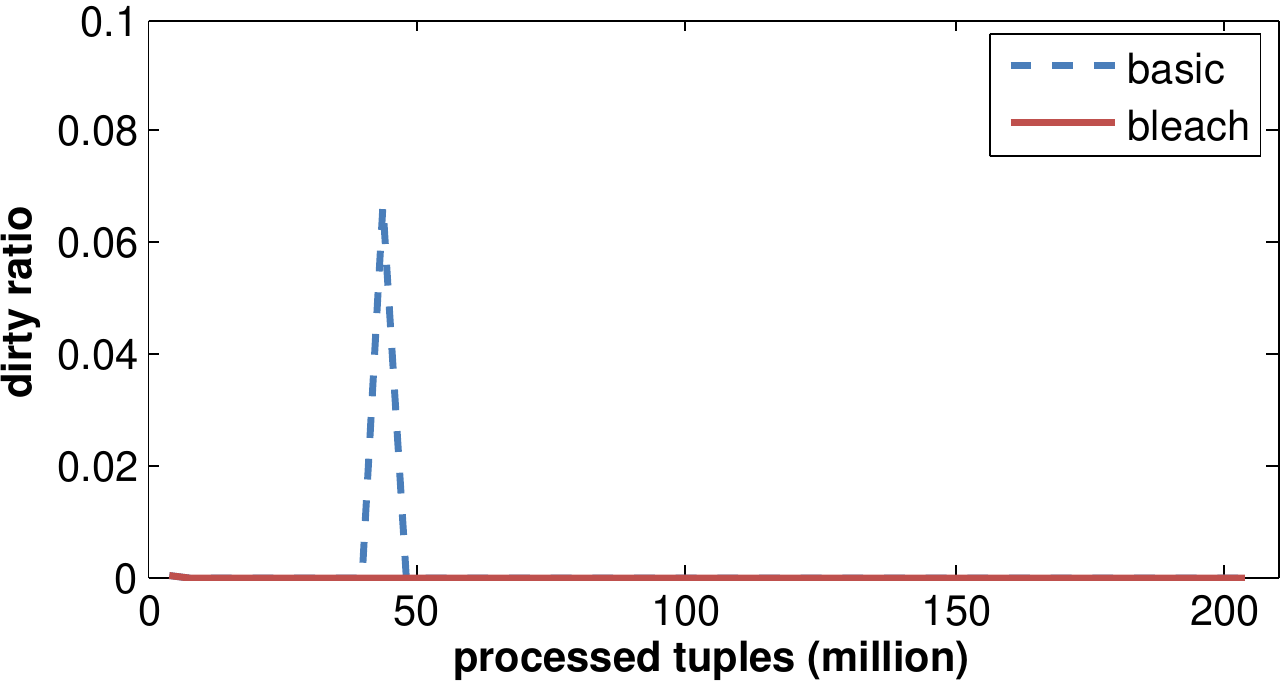}\label{fig:exp2_a_r4}}
	\subfigure[r5]{\includegraphics[width=0.32\linewidth]{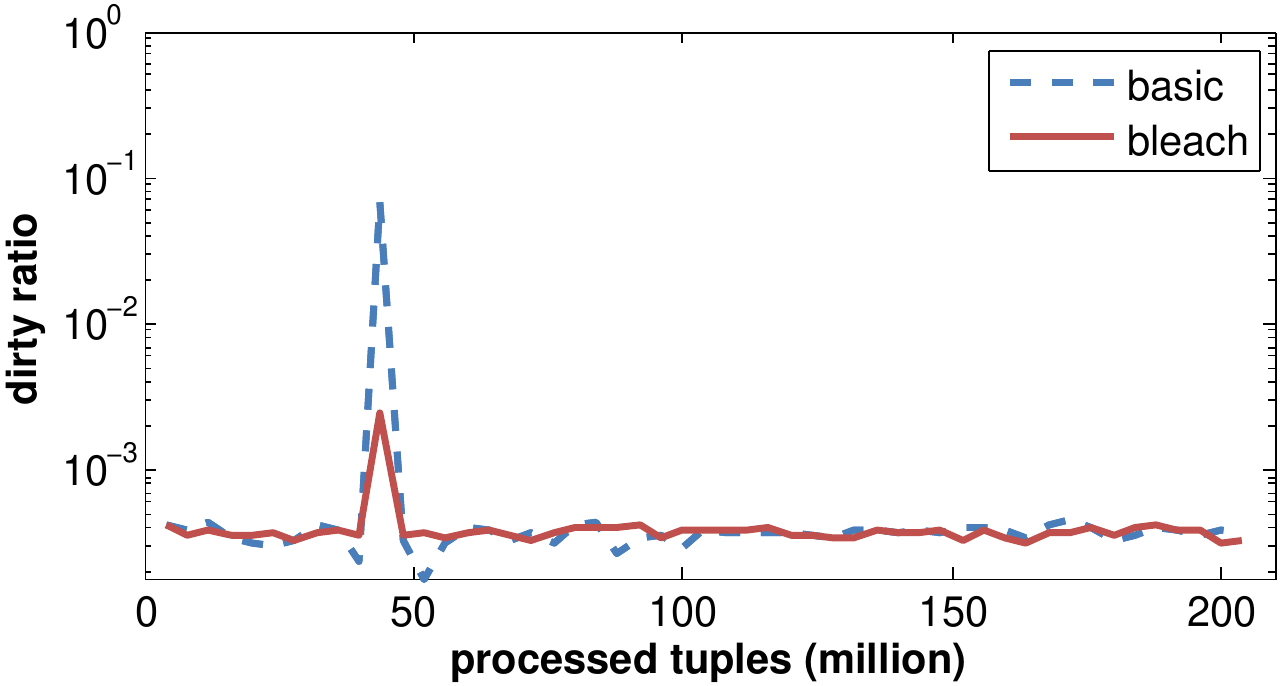}\label{fig:exp2_a_r5}}
	\caption{cleaning accuracy of two windowing strategies}
	\label{fig:exp2_accuracy}
	\vspace{-0.5em}
\end{figure*}

As shown in Figure~\ref{fig:exp2_throughput} and Figure~\ref{fig:exp2_latency}, the two windowing strategies are essentially equivalent in terms of throughput and latency: this is good news, as it implies the requirement for \emph{cumulative super cells} is a negligible toll on performance. 

Next, we focus on a detailed view of the cleaning accuracy, which is shown in Figure~\ref{fig:exp2_accuracy}. Bleach windowing achieves superior cleaning accuracy: in general, the dirty ratio is one order of magnitude smaller than that of basic windowing. This advantage is kept also in presence of a 50\% dirty ratio spike in the input data. In particular, for rules $r_3$ and $r_4$, Bleach windowing achieves 0\% dirty ratio, irrespectively of the dirty ratio spike.

Overall, Bleach windowing reveals that keeping state from past windows can indeed dramatically improve cleaning accuracy, with little to no performance overhead.

\subsection{Dynamic Rule Management}
\label{ssec:eva_3}

\begin{figure*}
\centering
\subfigure[Throughput\label{fig:exp3_throughput}]{\includegraphics[scale=0.44]{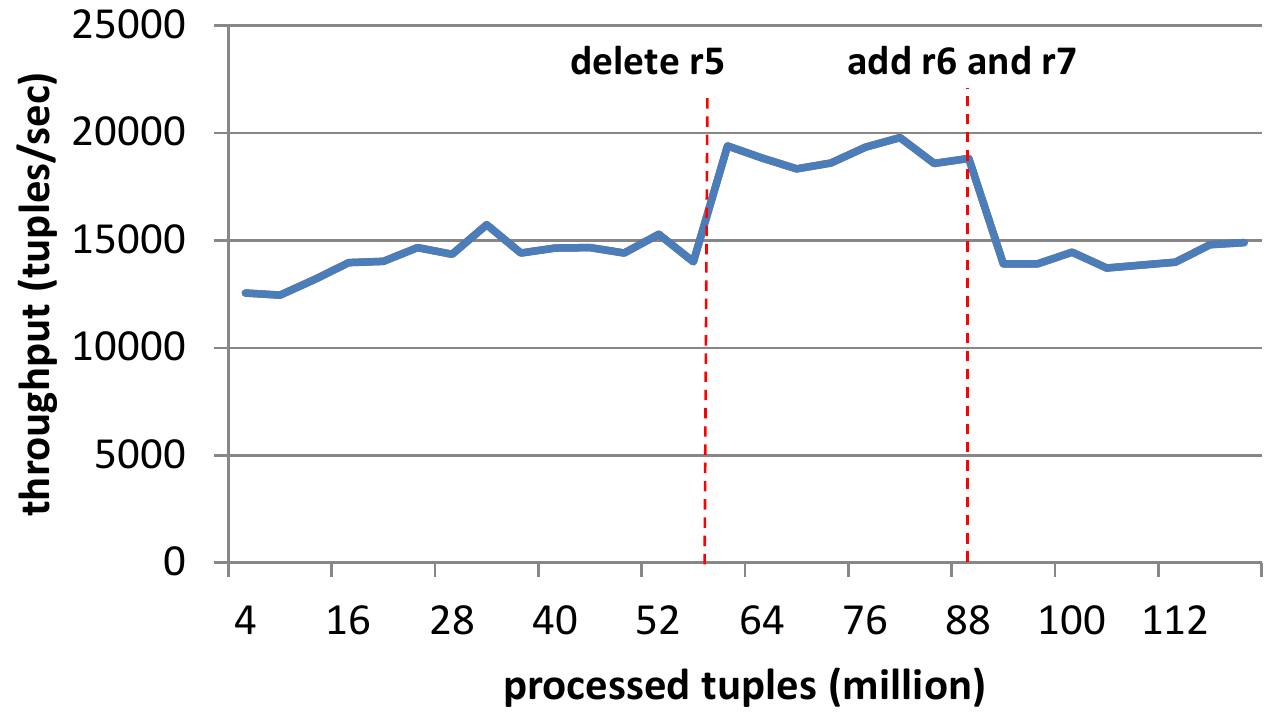}}
\subfigure[Latency in time\label{fig:exp3_latency}]{\includegraphics[scale=0.44]{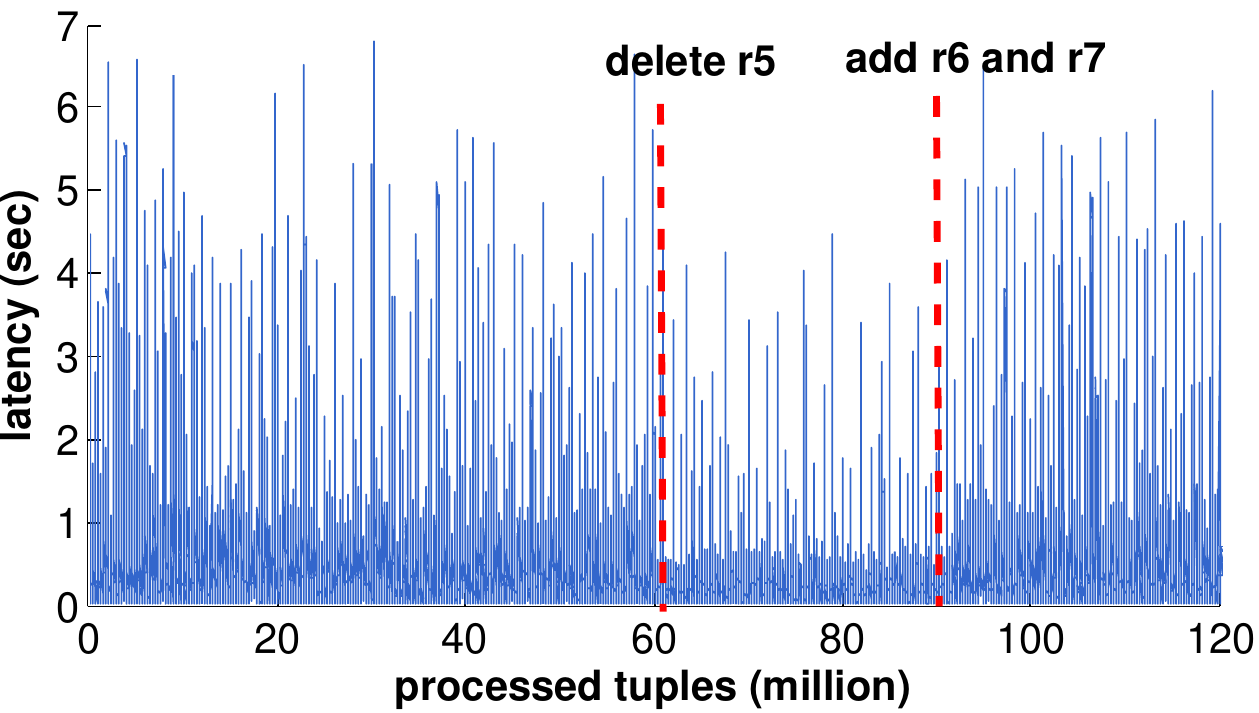}}
\subfigure[Latency CDF\label{fig:exp3_latency_cdf}]{\includegraphics[scale=0.44]{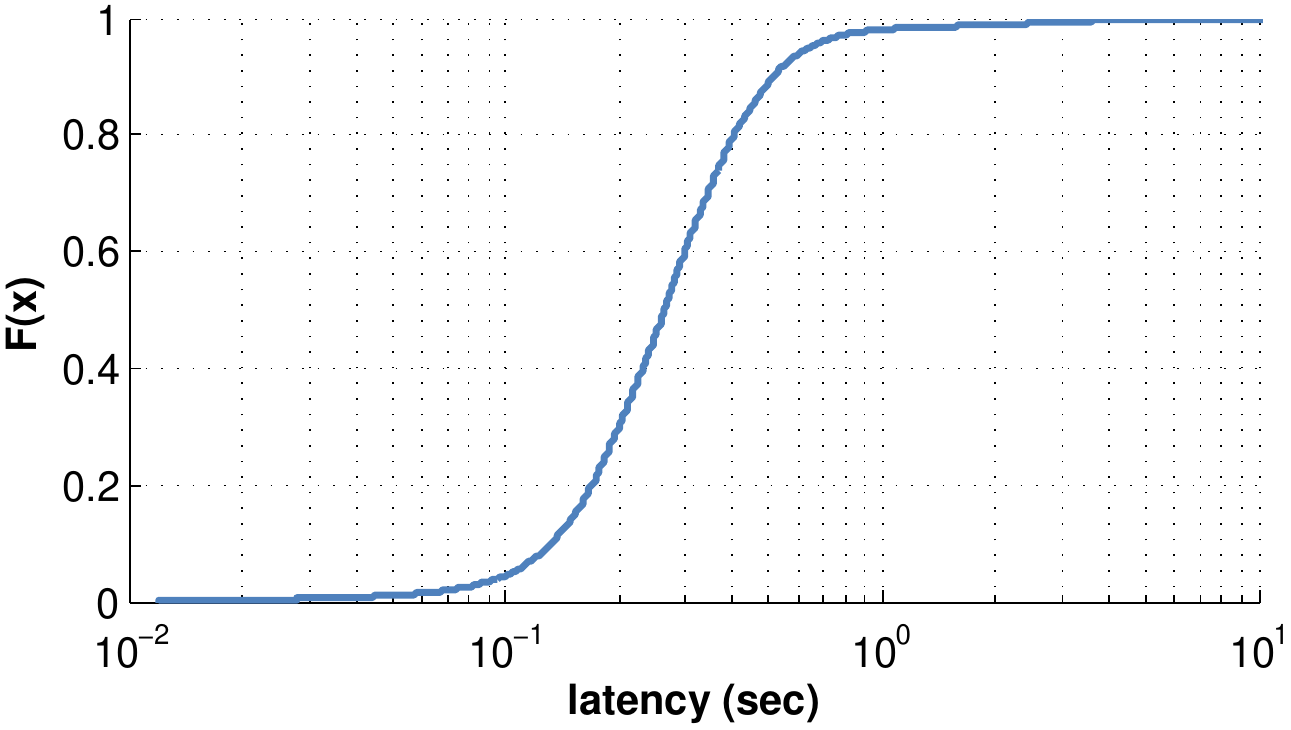}}

\caption{Bleach performance with dynamic rule management.}
\label{fig:dyn_rules}
\end{figure*}

Next, we study the performance of Bleach in presence of rule dynamics, as shown in Figure~\ref{fig:dyn_rules}. To do this, we initially use the same input data stream and rule set as in Section~\ref{ssec:exp1}. However, while Bleach is cleaning the input stream, we delete rule $r_5$ and add rule $r_6$ and $r_7$, as indicated in the Figure. In this experiment, we use Bleach windowing strategy and \emph{RW-dr} coordination.

Figure~\ref{fig:exp3_throughput} and Figure~\ref{fig:exp3_latency} show the evolution in time of throughput and latency, whereas Figure~\ref{fig:exp3_latency_cdf} gives the CDF of the processing latency.

Figure~\ref{fig:exp3_throughput} shows that rule dynamics can result in an increase in throughput. Indeed, removing $r_5$ (at the 60M tuple) implies that Bleach needs to manage fewer rules; in addition, $r_4$ becomes simpler to manage, as there are no more intersections with $r_5$. Similarly, Figure~\ref{fig:exp3_latency} shows that also latency decreases upon $r_5$ removal.
When rules $r_6$ and $r_7$ are added (at the 90 M tuple), the throughput drops and the latency grows, as Bleach has more rules to manage and because the new rules have intersecting attributes, requiring more work from RWs.

Figure~\ref{fig:exp3_latency_cdf}, shows the latency distribution computed from output tuple samples. While the average latency is roughly 320 ms, we notice a tail in the distribution, indicating that some (few) tuples experience latencies up to seconds. This has been observed across all our experiments, and is due to the sliding window mechanism, which imposes 
computationally demanding operations when updating the violation graph, resulting also
in rather low-level garbage collection problems.

Overall, we conclude that Bleach supports dynamic rule management seamlessly, with essentially no impact on performance, and no system restart required.

\subsection{Comparing Bleach to a Baseline Approach}
We conclude our evaluation with a comparative analysis of Bleach and a baseline approach, which follows the ideas discussed in Section~\ref{sec:intro}. 

To do so, we design and implement a new system that is based on the micro-batch streaming paradigm: essentially, such system buffers input data records and performs data cleaning periodically, as determined by a sliding window. Our implementation uses Apache Spark, and uses its Streaming API that supports all necessary operators.\footnote{To be precise, note that window processing in Spark Streaming is time-based and not tuple-based. For our experiment, this difference is negligible.} We refer to the baseline approach as micro-batch cleaning.

To demonstrate the performance of micro-batch cleaning and compare it to Bleach, we perform a series of experiments whereby we increase the sliding window size. We use the same stream data input from our previous experiments, but only use a single rule, $r_0$. Here, we focus on performance analysis expressed in terms of latency and dirty ratio, thus we feed the input stream at a constant throughput of 15000 tuples/second.

\begin{figure}
	\centering
	{\includegraphics[width=0.8\linewidth]{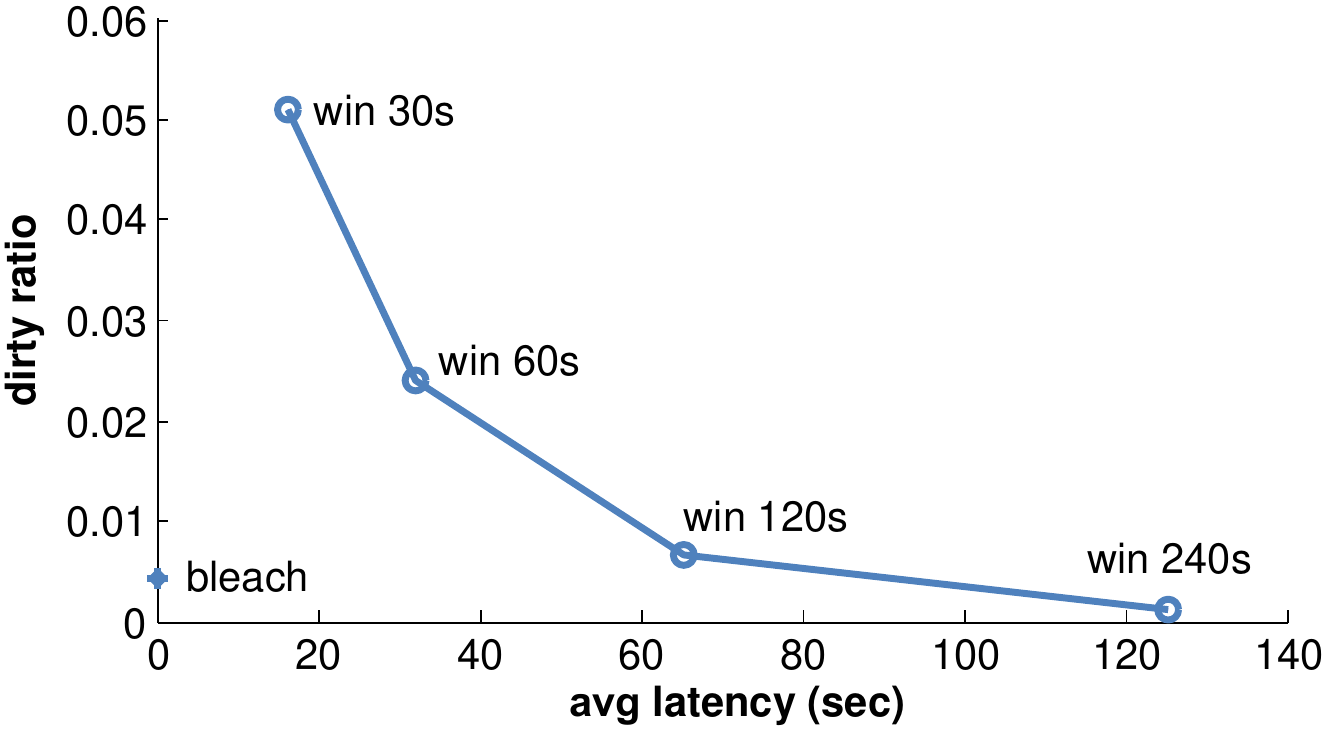}}
	\caption{micro-batch cleaning result}
	\label{fig:microclean}
\end{figure}

Figure~\ref{fig:microclean} illustrates the performance of both systems.
As expected, for the micro-batch baseline approach, the average latency is proportional to the window size: larger sliding windows entail higher latencies. Indeed, since the data in the input stream is uniformly distributed, the average latency equals the sum of half of the window size and the average execution time for cleaning data in each window.

As for the cleaning accuracy, the larger the sliding window, the more opportunities the micro-batch system has to clean the input stream, hence a smaller output stream dirty ratio. In particular, we notice that to achieve the same cleaning accuracy as Bleach, micro-batch cleaning requires the sliding window to be larger than 120 seconds, which incurs in an average latency larger than 1 minute. Instead, in Bleach, the average latency is about 320 ms. 

We conclude that Bleach represents a valuable tool in the data cleaning landscape if real-time requirements must be met, while achieving high cleaning accuracy.

%% file: 07-relatedwork.tex
\section{Related work}
\label{sec:rel}

In recent years, data cleaning systems have flourished. Many approaches \cite{4221723,Kolahi:2009:AOR:1514894.1514901,Beskales:2010:SRF:1920841.1920870,6544847,Dallachiesa:2013:NCD:2463676.2465327,Khayyat:2015:BSB:2723372.2747646,Geerts:2013:LDF:2536360.2536363} tackle the problem of detecting and repairing dirty data based on predefined data quality rules. 
\cite{6544847} proposes a way to combine multiple rules together and to perform data cleaning work holistically. 
\cite{DBLP:conf/dasfaa/ChenTHS015} focuses on functional dependency violations in a horizontally partitioned database, aiming to minimize data shipment and parallel computation time.
NADEEF\cite{Dallachiesa:2013:NCD:2463676.2465327} is a extensible and generic data cleaning system and BigDansing\cite{Khayyat:2015:BSB:2723372.2747646} is a large-scale version of NADEEF, which executes data cleaning job in frameworks like Hadoop and Spark.
These approaches are effective when data is static.

\cite{6228094,6243140} provide incremental algorithms to detect errors in distributed data when data is updated, which is similar to violation detection in Bleach.
\cite{6816655} introduces a continuous data cleaning framework that can be applied to evolving data and rules driven by a classifier.
These works focus on cleaning data stored in a data warehouse by batch processing, which achieves high accuracy but suffers high latency.

In contrast, stream processing requires to be real-time, a challenge that has drawn increasing attention from researchers \cite{43864,fernandez2015liquid,lin2015scalable,Elseidy:2014:SAO:2732279.2732281}. Nevertheless, stream data cleaning approaches are still in their infancy. 
Some works\cite{DBLP:conf/sigmod/SongZWY15,Zhao:2012:MAR:2396761.2396871,Jeffery:CSD-05-1413} focus on RFID or sensor stream data cleaning where the data is a sequence of numerical values. These works achieve data cleaning by operations like smoothing or deleting outliers.
Instead, Bleach focuses on more general cases where data can be both numerical and categorical.
In Spark Streaming~\cite{sparkstreamcleaning}, a data stream can be cleaned by joining it with precomputed information. However, precomputed information is not always available nor sufficient for accurate data cleaning. 
To the best of our knowledge, Bleach is the first stream data cleaning system based on data quality rules providing both high accuracy and low latency.

There are many other research work about data cleaning. For example, \cite{Wang:2015:CDA:2723372.2723739} and \cite{Chu:2015:KDC:2723372.2749431} are about how to perform data cleaning via knowledge base and crowdsourcing. BART \cite{arocena2015messing} is an dirty data generator for evaluating data-cleaning algorithms. \cite{abedjan2015temporal} studies the problem of temporal rules discovery for dirty web data. All such works are orthogonal to ours.

%% file: 08-futurework.tex
\section{Conclusion}
\label{sec:con}

This work introduced Bleach, a novel stream data cleaning system, that aims at efficient and accurate data cleaning under real-time constraints.

First, we have introduced the design goals and the related challenges underlying Bleach, showing that stream data cleaning is far from being a trivial problem. Then we have illustrated the Bleach system design, focusing both on data quality -- we have introduced dynamic rule sets, and a stateful approach to windowing -- and on systems aspects -- we have addressed problems related to data partitioning and coordination, which are required by the distributed nature of Bleach. We also have provided a series of optimizations to improve system performance, by using compact and efficient data structures, and by reducing the messaging overhead.

Finally, we have evaluated a prototype implementation of Bleach: our experiments showed Bleach achieves low-latency and high cleaning accuracy, while absorbing a dirty data stream, despite rule dynamics. We also have compared Bleach to a baseline system built on the micro-batch paradigm, and explained Bleach superior performance.

Our plan for future works is to support a more varied rule set and to explore alternative repair algorithms, that might require revisiting the inner data structures we use in Bleach.